\documentclass[acmtog,nonacm]{acmart}

\usepackage{booktabs} %

\usepackage{graphicx}
\usepackage{amsfonts}
\usepackage{amsmath}
\usepackage{pifont}
\usepackage{booktabs}
\usepackage{xcolor}
\usepackage{multirow}
\usepackage[normalem]{ulem}
\usepackage[super]{nth}
\usepackage{cleveref}
\usepackage{colortbl}

\newif\ifdraft
 \drafttrue
\ifdraft
\newcommand{\myc}[1]{{\color{blue}\textbf{Mingyi:} #1}}
\newcommand{\xlc}[1]{{\color{magenta}\textbf{Xuelin:} #1}}
\newcommand{\tkc}[1]{{\color{cyan}\textbf{Taku:} #1}}
\newcommand{\ruic}[1]{{\color{green}\textbf{Rui:} #1}}

\else
\newcommand{\myc}[1]{}
\newcommand{\xlc}[1]{}
\newcommand{\tkc}[1]{}
\newcommand{\ruic}[1]{}

\fi

\usepackage{xspace}

\newcommand{\keep}[1]{}
\newcommand{\old}[1]{}

\usepackage{url}
\usepackage{hyperref}

\newcommand{\cmark}{\ding{51}}
\newcommand{\xmark}{\ding{55}}

\newcommand{\nbframepred}{ F }
\newcommand{\nbframepast}{ P }

\newcommand{\sample}{ \boldsymbol{x} }

\newcommand{\pastmotion}{ \boldsymbol{c}_p }

\newcommand{\shapevec}{ \boldsymbol{\beta} }
\newcommand{\futuretraj}{ \boldsymbol{c}_{ft}}

\newcommand{\cond}{ \boldsymbol{c} }

\newcommand{\loss}{ \mathcal{L} }

\newcommand{\stride}{ s }

\usepackage[ruled]{algorithm2e} %
\usepackage{subcaption}

\newcommand{\name}{MotionPersona\xspace}

\acmJournal{TOG}

\begin{document}
\title[\name]{\name: Real-Time Locomotion Control across Personas, Bodies, and Styles}

\author{Mingyi Shi}
\email{irubbly@gmail.com}
\orcid{0000-0002-5180-600X}
\affiliation{\institution{The University of Hong Kong}\city{Hong Kong}\country{China}}
\author{Wei Liu}
\email{wloki.liu@gmail.com}
\affiliation{\institution{The University of Hong Kong}\city{Hong Kong}\country{China}}
\author{Jidong Mei}
\email{jidong_mei@connect.hku.hk}
\affiliation{\institution{The University of Hong Kong}\city{Hong Kong}\country{China}}
\author{Wangpok Tse}
\email{crazytse@connect.hku.hk}
\affiliation{\institution{The University of Hong Kong}\city{Hong Kong}\country{China}}
\author{Rui Chen}
\email{riorui@foxmail.com}
\orcid{0009-0003-7122-5207}
\affiliation{\institution{Hong Kong University of Science and Technology}\city{Hong Kong}\country{China}}
\author{Xuelin Chen}
\authornote{Corresponding authors.}
\email{xuelin.chen.3d@gmail.com}
\orcid{0009-0007-0158-9469}
\affiliation{\institution{Adobe}\city{London}\country{United Kingdom}}
\author{Taku Komura}
\authornotemark[1]
\email{taku@cs.hku.hk}
\orcid{0000-0002-2729-5860}
\affiliation{\institution{The University of Hong Kong}\city{Hong Kong}\country{China}}
\renewcommand{\shortauthors}{Shi et al.}

\begin{abstract}

We present MotionPersona, a generative framework for character-aware locomotion control, in which the motion produced for a command depends on the captured persona, the body shape, and the style of the character. Unlike style, which one performer can vary at will, persona and body shape are coupled in capture: each performer is observed in only one body. The captured data therefore cannot uniquely determine which motion characteristics should follow the persona and which should change with the body, leaving unseen persona--body combinations unconstrained.
We capture 48 performers, aged 5 to 68, under the same nine styles and seven commands, 44 of them with persona annotation. From this repeated-measures design, we identify two robust associations between body shape and gait. These measurements guide a cross-body specification of which motion characteristics should change and which should be preserved. We implement this specification through a physically informed retargeting pipeline, producing cross-body training data while penalizing penetration and foot skating.
On this cross-body data we train a single generative controller. A shape-aware VAE compresses each block of motion into a few latent tokens and renders them on a conditioned target body under explicit geometric supervision; over these tokens, a latent flow-matching prior generates persona- and style-conditioned motion in two sampling steps. The controller covers all captured personas, a wide family of SMPL-X target bodies, and nine styles in one model, and runs at 27 ms per block on two threads of a laptop CPU.
Finally, we verify the framework at every stage, following the same gait descriptors from captured to retargeted to generated motion and sweeping each axis in isolation. To our knowledge, this is the first real-time locomotion controller that carries part of a captured persona's performer-specific variation across independently selected body shapes and styles.

\end{abstract}


\begin{teaserfigure}
    \centering
     \includegraphics[width=\textwidth]{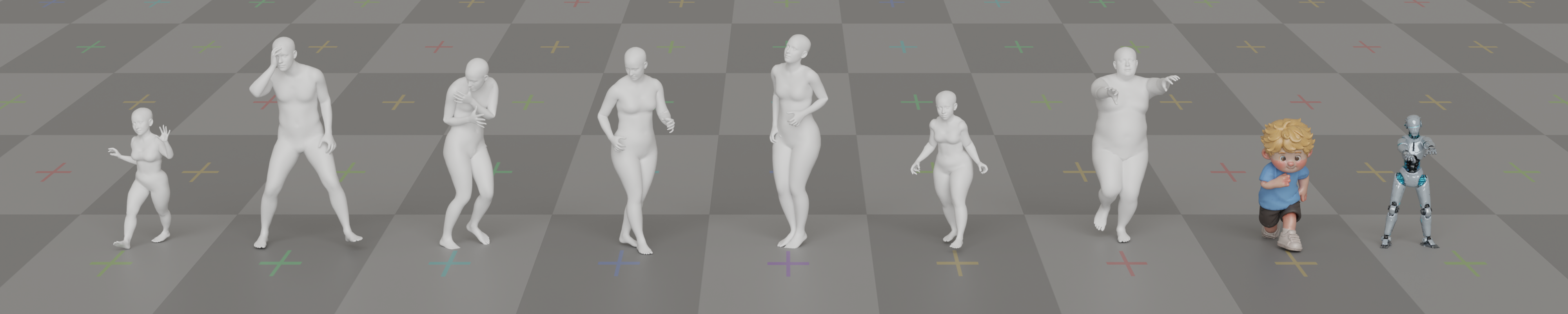}
     \caption{
       We present a generative character-aware locomotion controller whose three controls specify a captured motion persona (who is moving), a target body (which body carries the motion), and a performed style (how the character is moving) under varying locomotion commands.
       Within the captured catalog and target-body family, one shared model recombines the three controls in real time.
       Code and data will be fully released.
     }
     \label{fig:teaser}
\end{teaserfigure}

\maketitle

\section{Introduction}
\label{sec:intro}

Real-time locomotion controllers turn user input into animation, but the resulting motion should depend not only on the control signal, but also on who is moving, the body through which the motion is expressed, and how the character chooses to move. 
We refer to these three factors as \emph{persona}, \emph{body shape}, and \emph{style}, respectively.
Production pipelines commonly represent comparable distinctions with separate assets.
Each character receives its own animation set, driven by a state machine or motion matching~\cite{kovarMotionGraphs2002,MotionMatch,holden2020learned}; a new body is accommodated by retargeting, foot IK, and stride warping; and a new style is layered on top.
This separation works, but its authored cost grows with the number of characters, and every retargeting pipeline must choose which properties of the source motion to preserve when the body changes.
Sharing one animation set reduces this cost but makes different characters walk alike, while bodies far from the source require additional adaptation.

Learning-based controllers promise to share capacity across these assets.
Holden et al.~\shortcite{holdenLearningMotionManifolds2015,holdenDeepLearningFramework2016} learned motion manifolds in which latent editing produces new motion and transfers style, and later work separates content from style across unpaired clips~\cite{abermanUnpairedMotionStyle2020}.
Character-aware control, however, requires more than concatenating additional conditions.
When one control changes while the command and the other two remain fixed, the generated motion should move toward the measured target of that control and preserve the quantities assigned to the others.
Existing real-time controllers meet this requirement for style alone; to our knowledge, none can hold a persona fixed while the body changes, let alone set all three controls independently at interactive rates.
Our persona axis is deliberately closed: it covers 44 captured motion identities, each identified by a performer ID and described by three fixed attributes.
This raises two challenges.

First, no existing dataset covers many performers, each in many styles, under the same commands: locomotion datasets built for controllers come from one or a few performers~\cite{100style,pfnn,harveyRobustMotionInbetweening2020}, and large collections such as AMASS and HumanML3D~\cite{AMASSICCV2019,humanml3d} offer breadth of content rather than repeated styles across people.
Standard capture also observes each performer identity in one body.
It therefore constrains motion at the observed persona--body pairs but not the counterfactual cells in which the same persona occupies another body.
Nor is ``the same motion on another body'' unique without a preservation rule: matching timing, joint rotations, normalized steps, and footprints gives different answers as soon as body proportions differ.
Geometric retargeting commits to one such answer; when used as training supervision, it imposes rather than discovers the intended body counterfactual.

Second, most real-time controllers are built for one body~\cite{pfnn,zhang2018mode,starke2022deepphase,holden2020learned,camdm}.
CAMDM~\cite{camdm} showed that diffusion can drive a character in real time with style as a condition, but within the closed set of 100STYLE~\cite{100style}: one performer, one body, one hundred styles.
For many personas in many bodies, the challenge is to share capacity without averaging away their motion differences.
It is not clear how large such a model must be, or whether it can still run in real time.

Our approach makes the missing counterfactual explicit and selects one construction that can be measured and tested.
We proceed in five steps: capture, measure, construct, unify, and verify.
We first capture 48 performers, aged 5 to 68, each performing the same nine styles under the same seven locomotion commands.
This repeated design separates the style shared across performers from the stable performer effect and the performer--style interaction; it does not by itself identify a causal body effect.
We assign only robust, age-controlled shape associations to body; other stable variation remains with persona.
Two associations qualify: forward and side trunk lean increase with hip width.
Cadence and several normalized gait descriptors show no reliable shape association at the resolution of this study, so our construction treats them as preservation targets.
Cross-body retargeting then instantiates this specification, combining the two lean associations as soft targets with preservation, collision, contact, and quasi-static feasibility penalties.
Applied across bodies from 1.05 to 1.95\,m tall, it produces \name-X, an explicit persona--body cross design under this chosen definition.

To unify the three controls in a single model, we train one block-autoregressive generative controller on \name-X.
A shape-aware VAE encodes 45 future frames into nine latent tokens and decodes them together with the previous boundary frames and the target shape.
A latent flow-matching prior generates those tokens from motion history, trajectory, and all three controls in two sampling steps.
This division of responsibility is an architectural bias toward target-body-aware reconstruction in the codec and conditional motion selection in the prior.
The compact two-step prior is designed for a real-time block budget, and the controller runs at 27\,ms per block on two threads of a laptop CPU, one block per twelve committed frames.

To verify the pipeline, we follow the same gait descriptors from capture through retargeting to generation, and test each control while holding the other two fixed.
Our contributions are:

\begin{itemize}
\item \textbf{Problem formulation.}
We formulate locomotion control along three independently selectable axes, namely persona, body shape, and style, and identify a fundamental ambiguity in one-body-per-performer capture: unseen persona--body combinations are unconstrained by the training data.
\item \textbf{Capture, measurement, and the cross design.}
A repeated grid of 44 annotated performers measures stable performer effects and robust shape associations; a weighted retargeting objective instantiates the selected associations and preservation targets as \name-X.
\item \textbf{A unified real-time generative controller.}
A shape-aware motion codec and a two-step latent flow-matching prior recombine captured persona IDs, their typed attributes, target bodies, and nine styles in one controller.
\item \textbf{A verification protocol for the three axes.}
Single-axis sweeps on all three axes, scored against the captured data, show that each control acts through its own input and where the controller falls short of the data's spread; a decoder-versus-prior decomposition of the shape effect, and capacity and baseline studies, locate the design decisions that matter.
\end{itemize}

\section{Related Work}
\label{sec:related}

Our work connects motion retargeting and shape-conditioned generation, real-time locomotion control, and multi-performer datasets.
We review how each treats character variation.

\paragraph{Retargeting and shape-conditioned motion}
Motion retargeting moves a clip onto a new body and must decide which properties of that motion to preserve.
Kinematic methods change as little as possible: Gleicher~\shortcite{gleicher1998retargetting} solves for the smallest edit that keeps the feet planted, Choi and Ko~\shortcite{choi2000online} do the same online, frame by frame, and later pipelines automate the transfer onto new characters~\cite{feng2012automating}.
Physical methods add the dynamics of the target body: Tak and Ko~\shortcite{tak2005physically} filter the motion for balance, and Al Borno et al.~\shortcite{alborno2018robust} retarget through simulation onto realistic body shapes.
Contact- and surface-aware methods preserve relations within the skinned body or with other characters and objects~\cite{basset2019contact,jin2018aura,villegas2021contact,lakshmipathy2025kinematic,cheynel2025reconform}.
Learned methods replace the solver with a network trained across skeletons, often from unpaired data~\cite{delhaisse2017transfer,villegas2018neural,lim2019pmnet,abermanSkeleton2020,lee2023same}.
These methods preserve selected source-motion properties while repairing incompatibilities with the target body, such as sliding feet or a hand penetrating a wider torso.

Shape-conditioned generators incorporate the target body into motion generation itself.
MOJO~\cite{zhang2021we} projects predicted marker motion to a parametric body, while HUMOS~\cite{tripathi2025humos} explicitly targets cross-body motion with a target-body-conditioned decoder, cycle consistency, and physical losses despite the absence of paired motions across identities.
These methods can produce plausible motion on a new body, but when identity and shape covary in the observations, the interpretation of an unobserved identity--body combination still depends on objectives and inductive bias.
We make that choice explicit through measurement-guided association and preservation targets, implement them in a physically informed retargeter, and trace them into a controller.

\paragraph{Real-time character controllers}
Motion graphs and motion matching~\cite{kovarMotionGraphs2002,MotionMatch} drive a character by retrieving captured clips, and learned motion matching~\cite{holden2020learned} compresses the search into a network.
Phase-functioned and mixture-of-experts networks~\cite{pfnn,zhang2018mode,starkeLocalMotionPhases2020,starke2022deepphase} and recurrent models~\cite{lee2018interactive} regress the next pose directly and give stable, low-latency control, but a regressor averages the variation in its training data.
Generative controllers model a distribution instead.
VAEs~\cite{ling2020character, shi2023phasemp}, GANs~\cite{shiobara2021human,wang2021scene,kundu2019bihmp,men2022gan}, normalizing flows~\cite{henter2020moglow}, and diffusion models~\cite{mdm,zhang2022motiondiffuse,yuan2023physdiff,chen2023executing,alexanderson2023listen} recover diversity and detail, and autoregressive diffusion controllers such as CAMDM~\cite{camdm} and AMDM~\cite{amdm} bring this to interactive rates, over a hybrid representation~\cite{zhao2026autoregressive} or in reactive form against a live partner~\cite{shi2024takestwo}.
Generating through a learned latent and a prior over it follows motion VAEs~\cite{ling2020character} and latent diffusion~\cite{rombach2022high,chen2023executing}.
Few-step latent generators~\cite{dai2024motionlcm,tevet2025closd,gou2025control,shi2026prior} push this form toward real time, including flow matching in a per-pose latent driven by control operators.
MotionBricks~\cite{wang2026motionbricks} instead scales a latent in-betweening backbone across a large corpus, with behavior authored through keyframes.
Our controller also uses a latent codec and a flow-matching prior, but it differs in what the decoder is told: the boundary frames of the previous block and the body on which it reconstructs motion.

Physics-based controllers~\cite{peng2018deepmimic,peng2021amp,won2022physics,yao2022controlvae,dou2023c,juravsky2024superpadl} learn policies in simulation, where a new body is a change of simulation parameters.
AdaptNet~\cite{xu2023adaptnet} adapts a policy to a new morphology or style, and Generative GaitNet~\cite{park2022generative} varies body proportions in a simulated musculoskeletal model.
We keep the controller kinematic and use physics offline, in the retargeter that builds its training data.

Style is the one character axis these controllers already carry.
100STYLE~\cite{100style} conditions a phase-based controller on a style label, CAMDM conditions diffusion on the same labels, and a long line of style transfer~\cite{hsu2005style,xia2015realtime,yumer2016spectral,holdenDeepLearningFramework2016,abermanUnpairedMotionStyle2020,dong2020adult2child,guo2024generative,zhong2024smoodi} moves a style from one clip to another.
PersonaBooth~\cite{kim2025personabooth} personalizes an offline text-to-motion generator from multiple reference motions and promotes a cohesive persona across action contents.
Our present controller instead addresses real-time recombination of a closed catalog of captured performer identities with body shape and style.

\paragraph{Multi-performer data}
Locomotion datasets built for controllers hold one or a few performers~\cite{pfnn,harveyRobustMotionInbetweening2020,abermanUnpairedMotionStyle2020,100style,hou2024causal}, so who is moving and what they are doing vary together.
Broad corpora such as AMASS~\cite{AMASSICCV2019}, HumanML3D~\cite{humanml3d}, BABEL~\cite{BABEL:CVPR:2021}, and Motion-X~\cite{lin2023motionx} offer breadth of content across many subjects, but not the same styles under the same commands repeated across performers.
Our grid repeats styles and commands across performers, allowing shared style, stable performer effects, and performer-specific interpretations of style to be estimated separately.

\section{Problem Formulation}
\label{sec:problem}

\subsection{Character-aware control}
\label{sec:axes}
A real-time controller predicts the next block of motion from the history and the desired trajectory.
We write the two together as the command $\cond = (\pastmotion, \futuretraj)$.
A character-aware controller adds three axes to the command.
The \emph{persona} $\pi_s$ specifies who is moving.
In this work, persona denotes a closed-set, performer-specific motion identity: the characteristic motion signature of one of 44 annotated performers that persists across commands and styles.
The \emph{body shape} $\shapevec$ specifies the geometry of the body carrying the motion.
The \emph{style} $y$ specifies how the persona moves for the current command, using one of nine captured style labels.
The controller samples the next motion block $\sample$ from the learned
\begin{equation}
p(\sample \mid \cond, \pi, \shapevec, y).
\label{eq:controller}
\end{equation}

\subsection{The single-axis test}
\label{sec:single_axis}
The single-axis test asks what the conditions in Equation~\ref{eq:controller} should do: change one axis of the three, hold the other two and the command, and the motion should change only in the specified way.
A change of body should move the trunk lean and the foot contacts, and leave cadence, normalized step length, knee range, and hand distance alone.
Which descriptors persona and style own is read from the grid's variance decomposition in Section~\ref{sec:grid_reveals}: a descriptor belongs to the factor with the largest share of its variance, and to both when their interaction is largest.
By that rule a change of persona should move jerk, contact ratio, cadence, and step length, a change of style should move pelvis bob, head bob, and hand reach, and the two share elbow angle, arm swing, and trunk lean; speed and sway, whose largest share is the performer's, are fixed by the command in the test and should move under neither.
These lists are the targets of our specification, not claims of causal independence.
For the body axis, a fixed command means a command in the body's own units.
Speed is cadence times normalized step length times leg length, so the same command asks a longer leg to walk proportionally faster while preserving cadence and normalized step length.

\begin{figure}[t]
  \centering
  \includegraphics[width=\columnwidth]{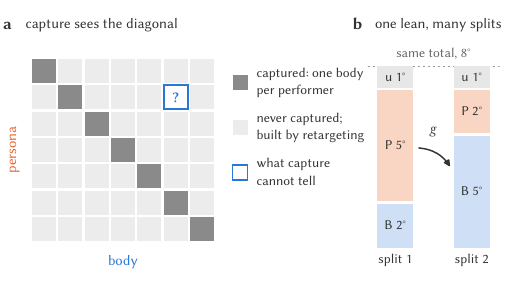}
  \caption{Why real capture binds persona to body.
  (a) Each captured performer brings one body, so capture fills only the diagonal of the persona $\times$ body grid (schematic; 44 performers in our data); every other cell, such as one persona on another body, is never observed, and the cross design of Section~\ref{sec:dataset} constructs it by retargeting.
  (b) Within one performer, an 8$^\circ$ forward lean splits into body, persona, and remainder as $B + P + u$ in many ways; moving any $g$ from $P$ to $B$ explains the clips equally well.}
  \label{fig:axes}
\end{figure}

\subsection{Problem: Real capture binds persona to body}
\label{sec:coupling}
Each performer is captured in one body (Figure~\ref{fig:axes}a).
For a gait descriptor $d$, such as cadence or forward lean, the repeated style--command grid supports the model
\begin{equation}
\begin{split}
d(s, y, v) = \mu &+ \mathrm{Style}(y) + \mathrm{Command}(v) \\
&+ \mathrm{Performer}(s) + \mathrm{Interaction}(s, y) + \varepsilon .
\end{split}
\label{eq:decomp}
\end{equation}
Style is shared across performers, Performer persists across styles, and Interaction describes how each performer plays a style.
Repeated observations across styles and commands let us estimate these terms.

But performer $s$ supplies only one persona--body pair $(\pi_s,\shapevec_s)$, leaving their contributions bundled:
\begin{equation}
\mathrm{Performer}(s) = B(\shapevec_s) + P(\pi_s) + u_s .
\label{eq:performer}
\end{equation}
Here $B$ is the contribution assigned to body shape, $P$ the contribution assigned to the closed-set motion persona, and $u_s$ the remainder.
Within one performer, the split between $B$ and $P$ is arbitrary (Figure~\ref{fig:axes}b): moving any amount $g$ from one to the other explains every observed clip equally well.
More generally, the capture observes the diagonal set $\{(\pi_s,\shapevec_s)\}$ but not its cross-product.
For any fitted predictor, one may add a function that is zero on every observed pair and arbitrary on an unobserved pair $(\pi_s,\shapevec_{s'})$; the training likelihood is unchanged, but the cross-body counterfactual is different.
The missing prediction is therefore not identified by captured data and would otherwise be selected silently by the model's inductive bias.
We instead write down a conservative specification: only shape-associated regularities that pass our prespecified tests are assigned to body; other stable variation remains with persona.
Section~\ref{sec:grid_reveals} gives the tests and their detection boundary; metadata-matched performers provide a proxy check, not observations of one persona in several bodies or a causal estimate.

\section{\name-X: Capture, Measurement, and Retargeting}
\label{sec:dataset}

We capture a repeated performer--style--command grid, use its measurements to specify which motion characteristics change across bodies, and construct \name-X through retargeting.

\subsection{Capture}

We captured 48 performers with a 29-camera VICON system in a $5\,\mathrm{m}\times5\,\mathrm{m}$ volume at 120\,Hz.
Forty-four of them (24 male, 20 female), aged 5 to 68, consented to persona annotation and supply all motion used in the analyses and \name-X; the other four contribute only target bodies to the cross design.
The bodies range from 105 to 189\,cm and from 16 to 90\,kg, and include five children and several performers over sixty.
Every performer was asked to perform the same nine styles, \emph{neutral}, \emph{angry}, \emph{depressed}, \emph{happy}, \emph{fear}, \emph{drunk}, and the locomotion variants \emph{swimming}, \emph{two-foot jump}, and \emph{big step}, under the same seven commands: forward, backward, and sideways walking; forward, backward, and sideways running; and transitions.
Demonstrations were provided, but performers interpreted each style in their own way, enabling measurement of both shared style and performer--style interaction.
The grid is nearly full: most performers cover nearly all 63 style $\times$ command cells, and the annotated grid holds 2,590 clips and about 33 hours of motion.

\subsection{Annotation}

Each clip carries three layers of annotation, illustrated in the appendix.

\paragraph{Persona, attributes, and body}
Age, gender, height, and weight come from a questionnaire; MoSh++~\cite{AMASSICCV2019} fits a 10-dimensional SMPL-X~\cite{SMPL-X:2019} shape vector $\shapevec$ that defines each performer's skeleton and mesh.
Each performer also chose one keyword on each of three dimensions, role or life stage (\emph{child}, \emph{student}, \emph{retiree}, \ldots), sociability (\emph{withdrawn} to \emph{gregarious}), and assertiveness (\emph{submissive} to \emph{dominant}); the keywords stay fixed across a performer's clips.
We refer to the latter two fields as affiliation and dominance in the model.
These fixed attributes accompany the unique performer ID.

\paragraph{Style}
Each clip carries a style label and its group (affective, physiological, or locomotion variant), varying posture and gait.

\paragraph{Command}
Each clip carries its recording command; the controller instead receives recent motion and the desired trajectory.

\subsection{Gait descriptors}

We compute 27 gait descriptors per clip, covering timing, space, limbs, and trunk, including cadence, step length, knee range, hand distance, and forward lean.
All use the pelvis-facing frame, with lengths normalized by the performer's leg length, hip width, or shoulder width to remove simple body scaling.
We reuse these descriptors at every stage; the appendix gives their definitions and estimators.

\subsection{Measured body effects}
\label{sec:grid_reveals}

\begin{figure}[t]
  \centering
  \includegraphics[width=\columnwidth]{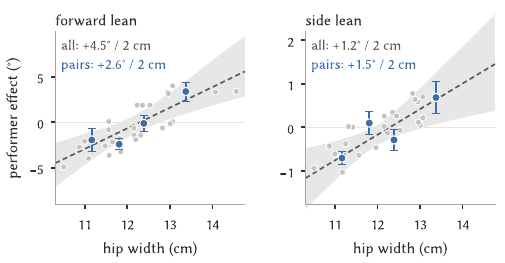}
  \caption{Wider hips go with more forward and side trunk lean, the two associations assigned to shape.
  Grey points are the adult performers' effects on trunk lean, the performer term of Equation~\ref{eq:decomp}, against hip width; blue points are the means of the four hip-width quartiles with standard errors.
  The dashed line and the band are the fit over all 39 adults and its bootstrap 95\% interval; the pair slope comes from the twelve metadata-matched adult pairs that share all three attributes but differ in body.}
  \label{fig:rules}
\end{figure}

\begin{figure*}[t]
  \centering
  \includegraphics[width=\textwidth]{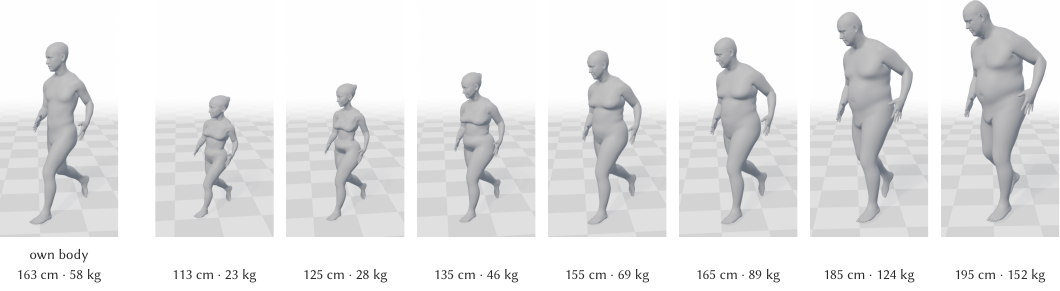}
  \caption{Cross-body retargeting.
  One frame of an angry forward walk on the source body (left) and seven target bodies of increasing height and girth (right: a captured child and six grid bodies), shown with a fixed camera rig and SMPL-X body masses.
  Stance widens, arms clear the torso and thighs, and trunk lean follows the assigned hip-width associations.
  On the widest body the plain copy penetrates by up to 14\,cm; the optimizer brings this to 1.6\,cm, and the mean forward lean runs from $-6.7^\circ$ on the child body to $+3.5^\circ$ on the widest.}
  \label{fig:retarget}
\end{figure*}

We report the findings that guide the cross-body specification; the appendix gives the full analysis, including every descriptor's variance shares, which set the ownership of Section~\ref{sec:single_axis}.

First, the performer effect is large and stable.
In Equation~\ref{eq:decomp}, performer effects exceed style effects by an order of magnitude for speed, jerk, and sway, with split-half reliability of 0.57--0.96 across styles, walking, and running, in line with perceptual evidence that gait carries identity~\cite{cutting1977recognizing,troje2002decomposing}.

Second, within the descriptors and statistical resolution of this study, body measures predict only a small part of the performer effect.
We regress each performer effect on nine body measures with leave-one-performer-out prediction, and assign an association to body only if it survives an age control, keeps its sign after partialling out age, appears with the same sign in all nine styles, and receives a directional check from the twelve metadata-matched adult pairs.
Two associations pass (Figure~\ref{fig:rules}): wider hips go with more forward trunk lean, 131$^\circ$ per metre of hip width within the pairs and 226$^\circ$ across the population, and with more side lean, 73 and 59$^\circ$ per metre; the 95\% intervals of the pair slopes, 41 to 249 and 19 to 135, contain the targets of 150 and 65 that the retargeter uses.
No other descriptor passes all tests: we detect no reliable body association with cadence, the apparent ones for normalized step length, knee range, foot clearance, and joint speed vanish under the age control, and hand distance scales with shoulder width; these become preservation targets of our construction, not universal invariants.
The five children form a strong but confounded signal, since the \emph{child} label, age, and small body coincide in this grid.
Overall, a variance-weighted shape predictor explains at most 7.5\% of the performer differences and 2\% of the total variance.

The three persona attributes do no better, with a leave-one-out $R^2$ of at most 0.18, so we leave the large and reliable remainder with the performer-specific persona ID.
The retargeter below implements the two passing associations and the preservation targets.

\subsection{Cross-body retargeting}
\label{sec:retarget}

The retargeter is our definition of the same motion on another body (Figure~\ref{fig:retarget}).
Given a captured clip and target body, it balances three sets of soft objectives: the shape associations assigned above, the selected preservation targets, and body-aware kinematic and quasi-static feasibility terms.
An explicit optimizer makes these terms and their residuals inspectable; the weighted objectives do not guarantee feasibility.
Applied to the 2,570 grid clips of at least 90 frames and to all 128 bodies, it yields \name-X: 328,960 retargeted realizations, about 4,200 hours from 33 source hours, in which persona and target body vary independently by construction.

\paragraph{The body}
A target shape $\shapevec$ defines the SMPL-X skeleton and a rest-pose mesh partitioned by skinning weights into fifteen rigid blocks, each with a signed-distance field, mass, and centre of mass.
Foot vertices define sole height, four support corners, and foot length.
We build 128 bodies once: the 48 captured performers, including the four without persona annotation, and 80 grid bodies that tile ten heights from 1.05 to 1.95\,m against eight girths, so that the cross design reaches beyond the performers we captured.

\paragraph{The optimizer}
The optimizer starts from the geometric copy: rotations copied, root path scaled by the ratio of leg lengths and root height by the ratio of hip heights, lowest toe on the floor.
The copy preserves cadence and normalized step length, which is what a fixed command means across bodies.
It is also what MotionBuilder's HumanIK retargeting produces on our capture: on 132 retargets across twelve performers its root-relative poses differ from the copy by 0.7\,cm and its stance skating equals the source's scaled by the leg ratio, so the copy stands in for the commercial tool wherever we need a geometric baseline.
Joint rotation corrections $\delta_t$ and root offsets $\boldsymbol\rho_t$ use twice-differentiable cubic B-splines with control points every three frames; target-body forward kinematics gives the corrected joint positions and block frames.
The objective is

\begin{equation}
E = \sum_i w_i\, E_i ,
\label{eq:retarget_obj}
\end{equation}
each term an average over the frames of the clip in centimetres, degrees, or radians, so that one set of weights serves every clip and every body.
The terms are, with their definitions and weights in the appendix:
\begin{itemize}
\item $E_{fid}$, $E_{root}$: fidelity, which pulls the rotation and root corrections back toward the copy;
\item $E_{pen}$: penalizes penetration between body blocks beyond a per-point allowance calibrated from both bodies;
\item $E_{ground}$, $E_{stance}$, $E_{slide}$, $E_{anchor}$: penalize feet below the floor, contact height, sliding during contact, and drift from the copy's contact anchors;
\item $E_{bal}$: penalizes a quasi-static zero-moment-point proxy outside the support region;
\item $E_{lim}$, $E_{smooth}$: penalize motion outside the source's observed joint range widened by 0.15\,rad, and nonsmooth corrections;
\item $E_{effort}$: applies a soft budget to a mass-weighted squared-acceleration proxy relative to the source;
\item $E_{lean}$: shifts mean forward and side lean by $k_f\,\Delta w$ and $k_s\,\Delta w$ for hip-width difference $\Delta w$, with $k_f = 150^\circ$ and $k_s = 65^\circ$ per metre, between the within-pair and population slopes.
\end{itemize}

One clip and one body cost about five minutes on a CPU; on twelve GPU shards the full run took about two days.
Every clip of \name-X is frame-aligned with its source and carries the persona ID and attributes of the source and the shape vector of the target.
The appendix reports the quality gates as residual distributions.

\paragraph{Check against real pairs}
Twelve adult pairs who share role, affiliation, and dominance but not body check the assigned associations against real people: each performer is retargeted into the other's body and compared with the other's own clips (Table~\ref{tab:pairs}).
Ours moves forward lean toward the other performer in 22 of 24 transfers and side lean in 19; forward lean moves by a median 1.9$^\circ$, against 1.4$^\circ$ assigned by the associations and 2.0$^\circ$ in the capture.
MotionBuilder, which copies joint rotations, moves neither and keeps the capture's floor penetration.
The pairs share performers, since three of the six cells hold three adults, and they also entered the selection of the associations, so this is a check of internal consistency, not an independent validation.

\begin{table}[t]
\centering
\caption{Ours moves the lean toward the real partner and removes floor penetration; MotionBuilder does neither.
Upper rows: medians over 24 transfers within twelve pairs; Capture is the real difference between the two performers, Assigned the part the associations predict from their hip widths, and Sign the fraction of transfers in which ours moves toward the partner.
Lower rows: 100 clips both retargeters produced on the same body.}
\scriptsize
\setlength{\tabcolsep}{4pt}
\begin{tabular}{l|cc|cc|c}
\toprule
& Capture & Assigned & MotionBuilder & Ours & Sign\\
\midrule
Forward lean, $^\circ$ & 2.0 & 1.4 & 0.0 & \textbf{1.9} & \textbf{0.92}\\
Side lean, $^\circ$ & 0.4 & 0.6 & 0.0 & \textbf{0.6} & \textbf{0.79}\\
Hand distance & 0.23 & -- & 0.00 & 0.03 & 0.54\\
Cadence (preservation target), Hz & 0.23 & -- & 0.00 & 0.02 & 0.54\\
\midrule
Penetration, \% of frames$\downarrow$ & 8.9 & 0 & 8.2 & \textbf{0.0} & --\\
Lowest toe, cm & $-4.1$ & $\geq 0$ & $-4.0$ & $\mathbf{-0.6}$ & --\\
\bottomrule
\end{tabular}
\label{tab:pairs}
\end{table}

\section{Generative Character-Aware Controller}
\label{sec:method}

\begin{figure*}[t]
  \centering
  \includegraphics[width=\textwidth]{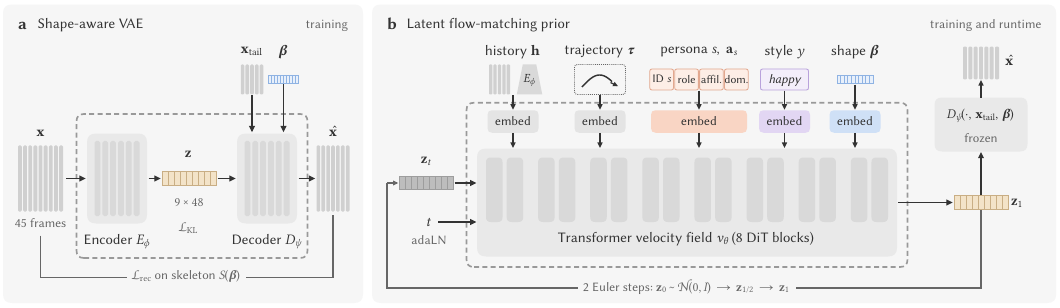}
  \caption{The controller: a shape-aware VAE realizes the motion on the target body, and a latent prior selects it.
  Dashed outlines mark the modules each stage trains.
  The shape-aware VAE decoder (a) realizes every block on the target body, continues the previous block, and is trained with explicit forward-kinematic and contact losses.
  The prior (b) generates the next latent tokens from the command, target shape, persona ID and typed attributes, and style, in two steps.
  }
  \label{fig:overview}
\end{figure*}

The retargeter produced \name-X, in which each captured persona has training realizations across the target-body family.
We now train one controller to share these combinations: a sampler for Equation~\ref{eq:controller} that generates 45 future frames from recent motion, a desired trajectory, and the three controls.

\subsection{Three-axis conditioning in a real-time budget}
\label{sec:why_two}

The controller must cover the persona catalog, target-body family, and nine styles within a real-time budget that also leaves room for rendering.
A block must be ready before the frames ahead of it run out; every millisecond saved can support more characters or a less powerful machine.
Limited capacity can average away subtle conditional effects, including the retargeter's few degrees of lean and centimetres of contact correction.
A one-stage pose-space controller shows it: it keeps under a fifth of the pelvis-bob difference between two performers (Section~\ref{sec:eval_capacity}).
Such averaging loses the distinctions that the cross-body training data was constructed to retain.
The challenge is to retain these differences without memorizing every combination as a separate distribution.

A VAE learns a compact motion representation and decodes it on the target body under forward-kinematic, contact, and boundary losses.
A flow-matching prior then learns to select motion in this codec space from the command and three controls.
Shape conditions both stages because morphology can affect motion selection as well as geometric realization.
This division is an architectural bias, not a strict factorization, and is tested through swaps and ablations.

\subsection{Architecture}
\label{sec:design}

\begin{table}[t]
\centering
\caption{Condition design and training dropout. Only history is dropped, enabling guidance in Equation~\ref{eq:cfg}.}
\scriptsize
\setlength{\tabcolsep}{3pt}
\begin{tabular}{@{}p{0.19\columnwidth}|p{0.37\columnwidth}|p{0.19\columnwidth}|p{0.10\columnwidth}@{}}
\toprule
Condition & Form & Enters & Drop\\
\midrule
History & last 5 frames & decoder & --\\
 & 1 token, codec encoder & prior & 15\%\\
Trajectory & 45 positions, 45 facings & prior & --\\
Persona ID $s$ & 44-way learned embedding & prior & --\\
Attributes $\mathbf a_s$ & three typed embeddings & prior & --\\
Style $y$ & label embedding & prior & --\\
Body $\shapevec$ & 10-d vector & decoder, prior & --\\
\bottomrule
\end{tabular}
\label{tab:conditions}
\end{table}

A block spans $\nbframepast + \nbframepred = 55$ frames at 30\,fps: $\nbframepast = 10$ history frames and $\nbframepred = 45$ future frames to generate.
Both stages read only the last five of the past frames; the raw-space baseline reads its default ten.
A frame holds the 23 joint rotations in the 6D representation~\cite{Zhou_2019_CVPR}, the root position, and two foot-contact flags, so the future part of a block is $\sample \in \mathbb{R}^{\nbframepred \times D}$.
Positions are relative to the last history frame: its root anchors both the block's root path and the desired trajectory.
This anchoring removes the character's place and heading in the world from every input, so a step learned in one part of the capture volume applies anywhere on the route.
As the rollout advances, committed generated frames become the history for the next prediction.

Figure~\ref{fig:overview} shows the computation within a block.
During training, the codec encodes the future motion into latent tokens and learns to reconstruct it on the target body.
At runtime, the prior supplies these tokens instead, and the decoder renders them as a continuation of the recent motion.

\paragraph{Conditions}
Table~\ref{tab:conditions} summarizes the inputs.
History enters the two stages in different forms.
The decoder receives the last five frames directly, providing the boundary from which the new block must continue.
The prior receives the same five frames as one token from the codec's encoder, so history and generated motion share a representation.
The prior's trajectory input comprises 45 root positions in metres and 45 facing directions relative to the last history frame, one token per frame in each stream.
Per-frame tokens let attention relate each generated token to the stretch of path it has to cover, which a single pooled trajectory vector would blur at turns.
For persona $\pi_s=(s,\mathbf a_s)$, a learned ID token identifies performer $s\in\{1,\ldots,44\}$, including performers who share the same attribute profile.
The ID is needed because the attributes are too coarse to identify a gait: they predict the descriptors with a leave-one-out $R^2$ of at most 0.18 (Section~\ref{sec:grid_reveals}).
Three separate embedding tables encode role, affiliation, and dominance, contributing one token each.
Typed tables distinguish, for example, \emph{moderate} affiliation from \emph{moderate} dominance.
The attributes remain attached to the selected ID, rather than acting as independent persona edits.
Style enters as a learned label embedding, as in CAMDM~\cite{camdm}.

\subsection{Shape-aware VAE}
\label{sec:vae}

The codec takes over everything geometric, so that the prior only has to choose the motion.
Forward kinematics on the target skeleton and foot contacts are losses on decoded frames, which the prior's objective in latent space cannot express; the codec is therefore where they are enforced.
The encoder $E_\phi$ groups every $\stride = 5$ frames into a token, yielding $n = 9$ tokens per block, then uses a six-layer transformer to predict a Gaussian posterior:
\begin{equation}
q_\phi(\mathbf z \mid \sample) = \mathcal N\big(\boldsymbol\mu_\phi(\sample),\ \mathrm{diag}\,\boldsymbol\sigma^2_\phi(\sample)\big), \qquad \mathbf z \in \mathbb R^{n \times d_z},\; d_z = 48 .
\label{eq:vae_enc}
\end{equation}
The decoder reconstructs motion on shape $\shapevec$, continuing the $n_{tail} = 5$ boundary frames $\sample_{tail}$:
\begin{equation}
\hat{\sample} = D_\psi(\mathbf z,\ \sample_{tail},\ \shapevec).
\label{eq:vae_dec}
\end{equation}
$D_\psi$ decodes one frame per output position rather than one patch per token: each token is expanded with $\stride$ learned frame queries, the 45 queries follow the shape, boundary, and latent tokens through a six-layer transformer, and a linear layer reads one frame from each.
A residual temporal head of three blocks, each a depthwise convolution over time (kernel 5) followed by a per-frame projection, then smooths the decoded frames across token boundaries with a receptive field of $\pm 6$ frames; its last projections are zero-initialized, so the decoder starts as the plain per-frame read-out.

Four choices determine how the codec reconstructs motion.
First, the decoder sees the target shape: without it, rollouts on extreme bodies penetrate the floor.
Second, it sees the boundary frames so that each block is rendered as their continuation.
Reconstruction looks similar with or without these frames; their benefit appears during generation, in the seam and contact quality.
Third, a token spans five frames rather than fifteen, because a longer span blurs when the feet touch the floor.
Fourth, the decoder must not read a token out as a block of frames.
A decoder that projects each token to its five frames with one linear layer leaves nothing to tie the last frame of one token to the first of the next: on the held-out takes of Section~\ref{sec:eval}, the jerk of its decoded toes at token boundaries is 2.2$\times$ the jerk between them (1.0 on the captured motion), the median jerk of a planted toe is 0.72 against the data's 0.51, and the flaw survives into every rollout as a visible roughness of the feet.
Per-frame queries with the temporal head bring the boundary ratio to 0.97 and the planted-toe jerk to 0.53 while halving penetration (0.5\% of foot frames); a three-frame token makes the roughness worse.

The training loss holds the VAE to the same geometry as a raw-space controller, on the conditioned body:
\begin{equation}
\loss_{\mathrm{VAE}} = \loss_{\mathrm{rec}}(\hat\sample, \sample;\ \shapevec) + \beta_{\mathrm{KL}}\ \mathrm{KL}\big(q_\phi(\mathbf z \mid \sample)\ \|\ \mathcal N(0, I)\big),
\label{eq:vae_loss}
\end{equation}
with the reconstruction loss
\begin{equation}
\loss_{\mathrm{rec}} = \loss_{\mathrm{pose}} + 5\,\loss_{\mathrm{vel}} + 0.5\,\loss_{\mathrm{pos}} + 5\,\loss_{\mathrm{pvel}} + 100\,\loss_{\mathrm{contact}} + \loss_{\mathrm{seam}} .
\label{eq:rec}
\end{equation}

The first two terms compare poses and their frame differences in feature space.
The next two compare joint positions $\mathbf p = \mathrm{FK}(\hat\sample, S(\shapevec))$ and velocities, always using the target skeleton $S(\shapevec)$ rather than an average body.
A foot that rests on the floor on an average body can be underground on a taller one; these losses measure the error on the body that will carry the motion.
The contact term penalizes decoded foot velocity during target contacts:

\begin{equation}
\loss_{\mathrm{contact}} = \frac{1}{\nbframepred}\sum_{f \in \{\mathrm{feet}\}}\ \sum_{t \in \mathcal C_f} \|\mathbf v_f(t)\|^2 ,
\label{eq:contact}
\end{equation}

where $\mathcal C_f$ holds the frames in which foot $f$ of the target motion moves by less than 1\,cm on the conditioned skeleton, and $\mathbf v_f$ is the velocity of the decoded foot; the sum is averaged over all future frames, so a block with more contacts carries more weight.

The seam term compares the first $W = 5$ decoded frames with the data.
We prepend the true history to the decoded block, keeping the past fixed while measuring position, velocity, and acceleration errors across the boundary:

\begin{equation}
\loss_{\mathrm{seam}} = \sum_{u \in \{\sample,\, \mathbf p\}}\ \sum_{k=0}^{2} w_k \sum_{t \in \mathcal W_k} \big\|\Delta^k \hat u_t - \Delta^k u_t\big\|^2 .
\label{eq:seam}
\end{equation}

Here $\Delta^k$ is the $k$-th frame difference.
The weights $w_k$ are 0.5, 0.5, and 0.25 for $k = 0, 1, 2$, respectively.
The windows $\mathcal W_k$ cover the $W$ frames after the boundary for positions and the differences reaching one frame back across it for velocities and accelerations.
We apply the loss to both pose features $\sample$ and world-space positions $\mathbf p$.
The KL weight is $\beta_{\mathrm{KL}} = 5 \times 10^{-4}$, with the per-element KL clamped at 0.1 nats and the weight annealed over the first fifth of training.
We report codec reconstruction separately from generated motion to distinguish the effects of compression from those of sampling.

\subsection{Latent flow-matching generative prior}
\label{sec:prior}

The second stage must select among the many motions that a command admits, while preserving the selected persona and style.
We use flow matching~\cite{lipman2023flow,liu2023rectified} to learn a velocity field from noise to motion tokens along straight interpolation paths, then integrate it in a few sampling steps.
The straight paths are what the real-time budget needs: a nearly straight path is integrated accurately in a few large steps, and each step is one pass of the network.
Its supervision is the latent-velocity target below, while explicit geometric losses remain in the codec.
Gou et al.~\shortcite{gou2025control} generate per-pose latents in four Euler steps with an MLP conditioned by control operators.
Our transformer instead generates a 45-frame block's nine tokens in two steps, with separate tokens for the conditions.
Separate tokens let each generated token attend to the conditions it depends on, the path near its own frames or the persona throughout, instead of reading one fused vector.

\paragraph{Objective and sampling}
The tokens are normalized once, with per-dimension statistics from a fixed batch at the start of training, $\tilde{\mathbf z} = (\mathbf z - \boldsymbol\mu_z)/\boldsymbol\sigma_z$.
With $\boldsymbol\epsilon \sim \mathcal N(0, I)$, $t$ drawn uniformly on $(0, 1)$, $\tilde{\mathbf z}_1$ a sample from the posterior of Equation~\ref{eq:vae_enc}, and $\mathbf z_t = (1 - t)\,\boldsymbol\epsilon + t\,\tilde{\mathbf z}_1$, the training objective is

\begin{equation}
\loss_{\mathrm{FM}} = \mathbb E_{t,\, \boldsymbol\epsilon}\ \big\| v_\theta(\mathbf z_t, t;\ \mathbf C) - (\tilde{\mathbf z}_1 - \boldsymbol\epsilon) \big\|^2 ,
\label{eq:fm}
\end{equation}

where $\mathbf C = (\mathbf h, \futuretraj, \pi, y, \shapevec)$ collects the history token $\mathbf h = E_\phi(\pastmotion)$ from the codec's encoder, the trajectory, and the persona, style, and shape tokens of Table~\ref{tab:conditions}.
The velocity network $v_\theta$ is a transformer of eight DiT blocks~\cite{peebles2023dit} over 106 tokens, the conditions followed by the noisy tokens.
Time $t$ enters through adaptive layer normalization rather than as a token.
At runtime the field is integrated from noise to data in two Euler steps, $\mathbf z_{t+\frac12} = \mathbf z_t + \tfrac12\, v_\theta(\mathbf z_t, t;\ \mathbf C)$, and decoded once on the conditioned body:

\begin{equation}
\hat\sample = D_\psi\big(\boldsymbol\sigma_z \mathbf z_1 + \boldsymbol\mu_z,\ \sample_{tail},\ \shapevec\big).
\label{eq:fm_sample}
\end{equation}

Two steps are empirically sufficient in the learned codec space.

\paragraph{The shape token}
The prior's shape token is not redundant with the decoder's.
The decoder can account for much of target-body geometry, but the latent motion is not constrained to be independent of shape.
Without the shape token, the prior plans steps that the decoder cannot land on the target body: foot skate and root tracking error both rise in our ablation.
The assigned lean association is also strongest when both stages know the target body.
This motivates co-conditioning: the decoder's geometric adaptation alone does not recover the full measured change in motion.

\paragraph{Guidance}
Persona, style, and body are controls that must hold in every block, whereas history is the one input that can conflict with a new command, for example a walk continuing into a requested run.
History is therefore the only condition dropped during training, and the only condition to which we apply guidance.
With $v_{\mathbf h}$ the prediction given history and $v_\varnothing$ the prediction with it dropped, as in 15\% of training, the sampler uses
\begin{equation}
v = v_\varnothing + w\,(v_{\mathbf h} - v_\varnothing),
\label{eq:cfg}
\end{equation}
where $w = 1$ is ordinary prediction at no extra cost, $w < 1$ loosens the influence of history, and $w > 1$ strengthens it.
Persona, style, and body are never dropped or guided.

\subsection{Training}
\label{sec:training}

The two stages train in turn on \name-X: the VAE on Equation~\ref{eq:vae_loss}, then the prior on Equation~\ref{eq:fm} in the token space of the frozen VAE.
Both sample 55-frame windows from the same \name-X training split.
The VAE has 17.2\,M parameters, 8.6\,M in the encoder and 8.6\,M in the decoder, and trains for 60 epochs in 11.7 hours on two RTX 4090 GPUs; the prior has 19\,M parameters and trains for 100 epochs in 18.4 hours on four.
Each epoch is 6,000 batches of 1,024 windows, with the learning rate annealed from $10^{-4}$ to $10^{-5}$; the appendix lists all hyperparameters.

The sampler is stratified by performer so that each identity contributes equally during training.
Left-right augmentation mirrors the skeleton together with the motion: SMPL-X is asymmetric, and mirroring motion alone shifts contact targets by 1--2\,cm on mirrored samples.
The held-out takes, the withheld persona $\times$ body cells, and the held-out bodies of Section~\ref{sec:eval} enter neither stage.

\subsection{Runtime}
\label{sec:runtime}

A block costs one history-encoder pass, two prior passes, and one decoder pass (Table~\ref{tab:runtime_cost}); encoding future motion is needed only in training.
The streamer commits three frames per block while input changes, so a new command, persona, or style enters the next block, and twelve once input is steady; the tentative remainder is cross-faded over two frames with the next block.
Cruising speed is proportional to leg length, as the retargeter's scaling implies, and scaled by the selected performer's pace.
The appendix details the streamer and the trajectory controller.

\begin{table}[t]
\centering
\caption{Block runtime. Steps are timed separately on two threads of an Apple M3 Max CPU and include individual call overheads; totals are measured end-to-end, not summed.}
\scriptsize
\begin{tabular}{l|r}
\toprule
Step & ms\\
\midrule
Featurization of the history and the trajectory & 0.3\\
History encoder, one token & 1.1\\
Prior, two Euler steps & 13.7\\
Decoder, per-frame queries and temporal head & 5.6\\
\midrule
Block, M3 Max CPU, two threads & 26.5\\
Block, RTX 5080 GPU & 9.9\\
\bottomrule
\end{tabular}
\label{tab:runtime_cost}
\end{table}

\section{Evaluation}
\label{sec:eval}

We evaluate whether \name supports persona, body shape, and style control while producing high-quality locomotion that follows a desired trajectory.

\subsection{Experimental setup}
\label{sec:eval_setup}

\paragraph{Evaluation setup}
A test case is one persona, one body, and one style.
The controller generates a one-minute rollout along a fixed trajectory, shown in the appendix, whose control sequence sets route, speed, and facing over time, under the runtime's steady-state schedule: two Euler steps per block, twelve frames committed per block after a two-frame cross-fade with the previous block, and the committed frames becoming the next block's history.
Every rollout starts from the same canonical history, a neutral walk on the test body, so the initialization cannot reveal the persona or the style; all methods receive the same inputs.
Route positions and speeds are scaled with leg length, so one command is fixed in body-relative units on every body.
Rollouts are evaluated without foot locking, and every stochastic number is the mean over five sampling seeds from one fixed checkpoint.

\begin{table*}[t]
\centering
\caption{Ours leads the baseline in control and in most quality columns on all four test sets, and its quality holds off the performer's body.
Own body: every persona in every style on its own body; cross-body: every persona in every style on training bodies it was paired with, on training bodies whose pairing was withheld, and on bodies training never saw.
Jerk is the median third difference of the toe positions in cm per frame$^3$ (the data value is measured on the reference takes), skate and seam in cm per frame, penetration in percent of frames, recovery and style accuracy as retrieval rates, with the held-out takes retrieved against each other as the data reference; spread kept exists on the own-body set only.
Own body + retargeter is the modular pipeline: our controller generates on the performer's own body, and our retargeter carries the result to the target body.
Bold marks the best method per test set.}
\scriptsize
\setlength{\tabcolsep}{3pt}
\begin{tabular}{ll|ccccccc|cc|cc}
\toprule
& & \multicolumn{7}{c|}{Quality} & \multicolumn{2}{c|}{Control} & \multicolumn{2}{c}{Spread kept}\\
& Test set & FPD pose$\downarrow$ & FPD contact$\downarrow$ & Jerk$\downarrow$ & Skate$\downarrow$ & IoU$\uparrow$ & Pen.\,\%$\downarrow$ & Seam$\downarrow$ & Rec.$\uparrow$ & Style$\uparrow$ & Dyn.$\uparrow$ & Amp.$\uparrow$\\
\midrule
\name-X (data) & own body & -- & 0.83 & 0.51 & 0.34 & 0.57 & 0.5 & -- & 0.889 & 0.80 & 1.00 & 1.00\\
\midrule
\multirow{4}{*}{CAMDM~\cite{camdm}} & own body & \textbf{0.30} & 1.90 & \textbf{0.65} & 0.60 & 0.49 & 49.0 & 3.52 & 0.323 & 0.32 & \textbf{0.43} & 0.55\\
& cross-body, seen & 0.28 & 3.28 & \textbf{0.51} & 0.53 & 0.50 & 21.8 & 3.10 & 0.257 & 0.51 & $\cdot$ & $\cdot$\\
& cross-body, withheld & 0.27 & 3.25 & \textbf{0.51} & 0.53 & 0.51 & 21.8 & 3.10 & 0.255 & 0.42 & $\cdot$ & $\cdot$\\
& held-out bodies & 0.28 & 3.42 & \textbf{0.51} & 0.53 & 0.50 & 23.9 & 3.11 & 0.256 & 0.40 & $\cdot$ & $\cdot$\\
\midrule
\multirow{3}{*}{Own body + retargeter} & cross-body, seen & 0.34 & 3.17 & 0.53 & 0.51 & 0.53 & 1.2 & 3.49 & 0.413 & 0.69 & $\cdot$ & $\cdot$\\
 & cross-body, withheld & 0.35 & 6.30 & 0.53 & 0.55 & 0.54 & 0.8 & 2.99 & 0.482 & 0.68 & $\cdot$ & $\cdot$\\
 & held-out bodies & 0.34 & 6.25 & 0.53 & 0.54 & 0.53 & 0.9 & \textbf{2.99} & 0.460 & 0.68 & $\cdot$ & $\cdot$\\
\midrule
\multirow{4}{*}{Ours} & own body & 0.32 & \textbf{1.70} & 0.66 & \textbf{0.49} & \textbf{0.54} & \textbf{1.0} & \textbf{3.45} & \textbf{0.632} & \textbf{0.74} & 0.42 & \textbf{0.70}\\
& cross-body, seen & \textbf{0.26} & \textbf{1.57} & \textbf{0.51} & \textbf{0.45} & \textbf{0.55} & \textbf{0.4} & \textbf{2.98} & \textbf{0.507} & \textbf{0.72} & $\cdot$ & $\cdot$\\
& cross-body, withheld & \textbf{0.26} & \textbf{1.56} & \textbf{0.51} & \textbf{0.45} & \textbf{0.56} & \textbf{0.4} & \textbf{2.98} & \textbf{0.532} & \textbf{0.71} & $\cdot$ & $\cdot$\\
& held-out bodies & \textbf{0.26} & \textbf{1.60} & \textbf{0.51} & \textbf{0.45} & \textbf{0.56} & \textbf{0.4} & 3.00 & \textbf{0.514} & \textbf{0.72} & $\cdot$ & $\cdot$\\
\bottomrule
\end{tabular}
\label{tab:main}
\end{table*}

\paragraph{Test sets}
We split the captured data by source take before windowing and retargeting, so all windows, retargeted versions, and mirrored variants of a take stay in one partition.
The own-body set puts each of the 44 personas in each of the nine styles on the performer's own body; it is what capture alone can check, and every number not marked otherwise is measured on it.
Three cross-body sets put every persona in every style on other bodies, drawn from sixteen bodies spaced over the hip-width range that also serve the single-axis sweeps: twelve training bodies and four of the fifteen bodies training never saw.
The seen set pairs each persona with training bodies it was paired with in training; the withheld set with training bodies whose pairing was withheld, one quarter of the personas on each of the twelve, chosen once before training; the held-out set with all fifteen held-out bodies.
The held-out bodies are grid bodies across the height and girth range, so every performer's own body is in training, and every method shares the partition.
The sets concern new bodies and pairings within the captured catalog, not unseen identities.
\paragraph{Baseline}
The baseline is CAMDM~\cite{camdm}, a one-stage diffusion model in pose space, in its default configuration of eight layers at width 256 and four sampling steps.
It receives the same performer ID, typed persona attributes, style, body shape, and trajectory as our method, the same training data, and the same body-aware geometric supervision.
It reads the last ten past frames as pose tokens, where our codec folds the last five into one history token; it has 6.9\,M parameters against our 17.2\,M codec and 19\,M prior, and its training settings are in the appendix.
\subsection{Metrics}
\label{sec:eval_metrics}

\paragraph{Motion quality}
Two Fr\'echet distances compare Gaussian fits on generated and held-out \name-X motion under the same persona, body, and style: the pose FPD on root-relative joint positions, facing-aligned and normalized by leg length, and the contact FPD on each foot's lowest sole-corner height and toe speed, standardized by their spread in the data, which sees where the feet are (the held-out takes score 0.83 on it).
One geometric detector serves every method and the data: a toe is in contact while slower than 2\,cm per frame within 5\,cm of the sole plane.
Skate is the toe speed in contact, in cm per frame; penetration the fraction of frames with a sole corner more than 1\,cm below the plane; IoU the agreement of a model's contact flags with the detector on its own geometry, and of the captured labels with it for the data; the seam the joint displacement across a block boundary, in cm per frame, against about 3.4 within a block.
Cadence and step length use the grid's autocorrelation estimator, which does not depend on detected contacts; the appendix gives the remaining definitions.
\paragraph{Control accuracy}
Persona recovery retrieves the nearest of the 44 identities from a rollout's gait descriptors, against every identity's reference takes on the same body, retargeted to it when needed; style accuracy does the same among the nine styles of the performer.
Each identity's held-out takes are split in two, one half the reference and the other retrieved against it, giving the data reference in Table~\ref{tab:main}.
\paragraph{Spread kept}
A ratio of means cannot see averaging: two performers with 2 and 6\,cm of pelvis bob, both generated at 4\,cm, give ratios of 2 and 0.67.
Spread kept is, per descriptor, the standard deviation across the 44 personas of the generated value over the same in their references, one when the differences survive and zero when they are averaged; we report its mean over the dynamics descriptors and over the amplitude descriptors, listed in the appendix.
\subsection{Motion quality and character control}
\label{sec:eval_control}

Table~\ref{tab:main} compares ours with the baseline on the four test sets.

\paragraph{Against the baseline}
On the own-body set ours skates 0.49\,cm per frame against 0.60 for the baseline and 0.34 in the data, with a contact consistency of 0.54 against 0.49; the two part most on penetration, 1.0\% of frames against 49\%.
The feet are as smooth as the baseline's: the median toe jerk is 0.66 against 0.65 on the own body and 0.51 for both on every cross-body set, where the data measures 0.51.
The contact FPD is 1.70 against 1.90; the pose FPD, 0.32 against 0.30, is the one quality column the baseline wins clearly.
Style accuracy is 0.74 for ours and 0.32 for the baseline, against 0.80 for the data; ours keeps 0.70 of the amplitude spread between performers against 0.55, and the two are level on the dynamics spread, 0.42 against 0.43.
Persona recovery, a 44-way retrieval on six posture and gait descriptors selected on the data's own halves, so that the data value is an optimistic reference, reaches 0.63 for ours and 0.32 for the baseline on the own body, against a chance rate of 0.023 and 0.89 for the data; off the own body the baseline falls to 0.26 while ours stays at 0.51 to 0.53, and its style accuracy to 0.40 to 0.51 while ours stays at 0.71 to 0.72.
\begin{figure}[t]
  \centering
  \includegraphics[width=\columnwidth]{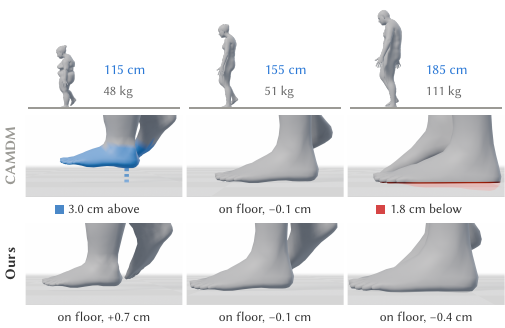}
  \caption{The baseline loses the floor at both ends of the height range; ours keeps it.
  One persona walks straight on three sweep bodies (thumbnails at a common scale); blue marks hovering feet, red the sole below the floor.
  Numbers are the median height of the lowest foot vertex above the floor over 19\,s; within $\pm1$\,cm counts as on the floor.}
  \label{fig:floor}
\end{figure}

\paragraph{Across bodies}
The three cross-body rows are the counterfactual capture cannot supply: the same persona on a training body it was paired with, on one it never was, and on a body training never saw.
For our controller the three rows agree to within 0.04 in every quality column, so a withheld pairing or a body training never saw costs nothing the metrics resolve relative to a seen pairing.
These sets test that the controller follows the specification of \name-X where training gave it no example; they cannot show how the performer would in fact move in another body, which the real pairs of Section~\ref{sec:retarget} check only in direction.
The baseline is a different controller on a different body: it penetrates a fifth of its frames, and its contact FPD rises from 1.90 on the own body to 3.3 to 3.4 off it, against 1.6 for ours, because its floor moves with the body.
Figure~\ref{fig:floor} shows it: on a 115\,cm body the baseline's feet hover 3\,cm above the floor, on a 185\,cm body they sink 1.8\,cm below it, and only near the middle of the height range does it stand where ours does on every body.
On the sixteen sweep bodies its contact ratio runs from 0.01 to 0.54 and its lowest toe from 0.1 to 4.5\,cm above the floor; ours stays within 0.48 to 0.61 and 1.3 to 2.5\,cm.

\paragraph{Against a modular pipeline}
The production alternative generates on the performer's own body and retargets the result.
With our retargeter in that role, the pipeline keeps the feet nearly out of the floor, 0.9\% of frames on the held-out bodies against 0.4\% for ours, but its contact FPD is 6.25 against 1.60, its recovery 0.46 against 0.51, and its style accuracy 0.68 against 0.72.
The retargeter also optimizes whole clips, about 0.14 GPU-seconds per second of motion, so it cannot follow a command that changes while the motion plays, whereas the controller commits each block in 27\,ms on a CPU.
\begin{figure*}[t]
  \centering
  \includegraphics[width=\textwidth]{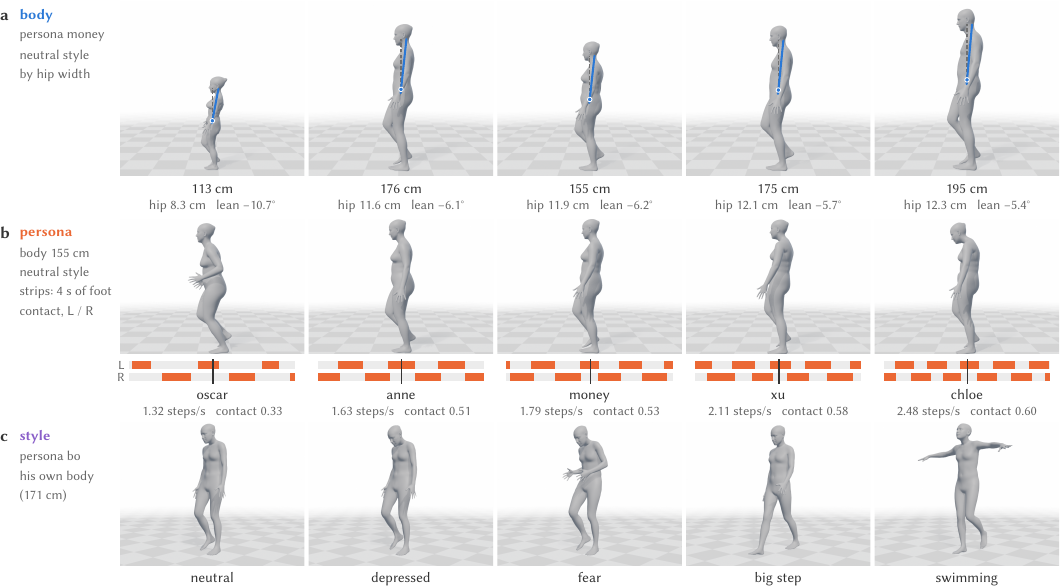}
  \caption{One control at a time, with the command fixed.
  (a) One persona on five bodies, ordered by hip width: the trunk (blue, against the grey vertical) leans back less as the hips widen.
  (b) Five personas on one body: the foot-contact strips, four seconds of the left and right foot, show five rhythms.
  (c) Five styles on one persona and body change the whole manner of moving.}
  \label{fig:gallery}
\end{figure*}

\begin{figure*}[t]
  \centering
  \includegraphics[width=\textwidth]{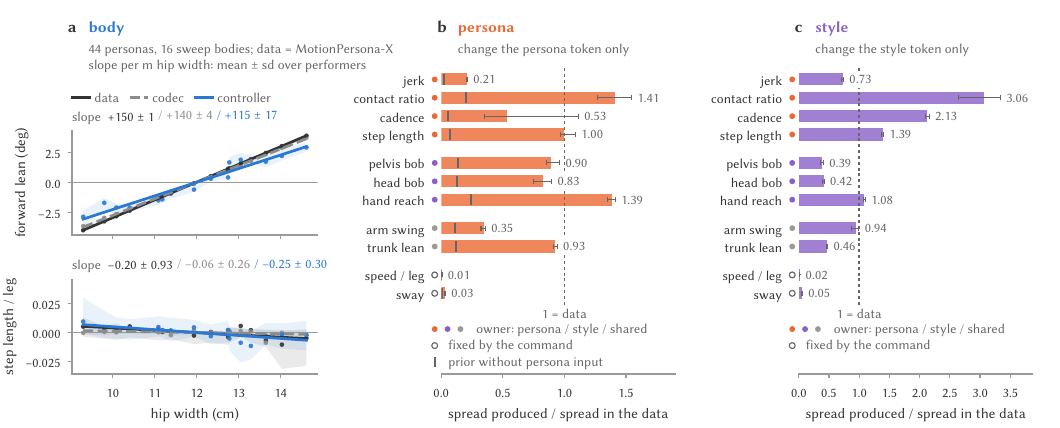}
  \caption{The single-axis test, scored against the data.
  Each column changes one control and holds the other two and the command.
  (a) Body: forward lean and step length against hip width for all 44 personas on the sixteen sweep bodies, in the held-out takes (black), their codec reconstruction (grey, dashed), and the controller's rollouts (blue); lines are means, bands the 10th to 90th percentile, numbers the mean and spread of the per-persona slopes per metre.
  (b, c) Persona and style: per descriptor, the spread the controller produces across the 44 personas or the nine styles, as a fraction of the data's spread on the same body with each command's mean removed; (c) averages over the personas; median over bodies with a bootstrap interval, 1 reproduces the data.
  Dots name the factor that owns the descriptor in the grid; open rows are fixed by the command and should stay near 0; ticks in (b) are a prior trained without a persona input.}
  \label{fig:sweeps}
\end{figure*}

\subsection{The single-axis test}
\label{sec:eval_axis}

The test is the one stated with the problem: change one control, hold the other two and the command, and the descriptors assigned to the control should move while the others stay.
It complements the table: a control can move the right descriptors toward the wrong target, as shuffled style labels would, and the recovery and accuracy columns rule that out.
The test shows which descriptors carry the change; Figure~\ref{fig:gallery} shows one example per control, and Figure~\ref{fig:sweeps} scores each descriptor against the data.
Every sweep uses the protocol above on the sixteen sweep bodies.

\paragraph{The body axis}
The shape sweep keeps one persona in the neutral style and moves the body across the sixteen bodies under the command scaled to each; Figure~\ref{fig:sweeps}a runs it for all 44 personas, at three points: the held-out takes retargeted to each body, their codec reconstruction, and the controller's rollouts.
The assigned lean should keep its slope from the first point to the last, and the preservation targets should stay flat.
Forward lean rises by 150$^\circ$ per metre of hip width in \name-X, the slope the retargeter was asked for, by 140$^\circ$ after the codec, and by 115$^\circ$ from the controller, with a spread of 17$^\circ$ across personas, so three quarters of the injected effect reaches the rollouts; the sweep persona of Table~\ref{tab:data} gives 105.
Side lean runs 64, 50, and 28$^\circ$ per metre: the codec keeps four fifths, the controller two fifths.
Step length and cadence stay flat at all three points.
The lost lean is shared between the persona input, which carries part of the source body's lean, and the prior's regularization: without the persona the prior reaches the target slope but loses the performers, and without dropout it overshoots, 1.28 of the target, and walks worse (Table~\ref{tab:ablation_prior}).
\paragraph{The persona and style axes}
The persona sweep keeps body, style, and command and moves the persona through all 44 identities on each of the sixteen bodies, in the neutral style; the style sweep keeps body, persona, and command and moves each of the 44 personas through the nine styles.
Neither has a frame-aligned reference, so each is scored against the data's own spread: per body and descriptor, the spread of the controller's means across the 44 personas, or the nine styles, divided by the spread across the 44 performers' held-out takes, or the nine styles, on that body, with each command's mean removed from the takes because every rollout follows one trajectory, and the median over bodies as the bar in Figure~\ref{fig:sweeps}b and c.
A bar at 1 reproduces the data; the open rows, speed and sway, are fixed by the trajectory and stay at 0.01 to 0.05 under both controls.
A prior trained without a persona input, the ticks in Figure~\ref{fig:sweeps}b, stays between 0.00 and 0.24 on every row, so the bars measure the persona input and not the body or the seed.
Under persona the controller reproduces the data's spread of step length, pelvis bob, head bob, and trunk lean, 1.0, 0.9, 0.8, and 0.9, exaggerates contact ratio and hand reach, 1.4, keeps half of cadence, 0.53 with a wide interval, and loses most of jerk and arm swing, 0.21 and 0.35.
Under style it reproduces hand reach and arm swing, 1.1 and 0.9, exaggerates the timing differences between styles, contact ratio 3.1, cadence 2.1, and step length 1.4, and flattens pelvis bob, head bob, and trunk lean to 0.39 to 0.46.
\paragraph{What the test shows}
We read the test against three criteria: attribution, whether a control moves the descriptors assigned to it beyond what the other inputs and the seed produce and leaves those the command fixes; fidelity, whether it moves them as far as the data; and exclusivity, whether descriptors assigned to another factor stay put.
Attribution holds: every persona bar exceeds the prior without a persona input, speed and sway stay near 0 under both controls, and the body keeps the sign and three quarters of the assigned forward lean while cadence and step length stay flat.
Fidelity is partial: of the descriptors the grid assigns most firmly, jerk to the performer and pelvis bob to the style, the controller keeps 0.21 and 0.39 of the data's spread, side lean keeps two fifths of its slope, and the style control exaggerates contact ratio and cadence 3.1 and 2.1 times.
A spread says how much the personas differ, not whether each differs in the right direction; the recovery column of Table~\ref{tab:main} tests that, at 0.51 to 0.53 off the own body against a chance rate of 0.023.
Exclusivity is not what the grid supports: ownership is the largest variance share, not the only one, so hand reach, which the style owns, also differs between performers in the data, and the controller reproduces that difference, 1.4 under persona, rather than suppressing it.
The controller therefore satisfies the specification in part: each control acts through its own input on the descriptors assigned to it, but it does not reproduce the data's structure in full, and the gap is the one the spread columns of Table~\ref{tab:main} report.
\subsection{What the cross design contributes}
\label{sec:eval_data}

\begin{table}[t]
\centering
\caption{What each step of the cross design contributes.
The same controller trained on the captured grid, on the geometric copies, on the optimizer's output without the lean targets, and on \name-X, with the data and the codec's reconstruction above.
Slopes per metre of hip width on the sweep bodies, targets under the headers (over all 44 personas, ours gives 115 and 28); penetration and preservation on the same bodies, recovery on the held-out bodies.
Bold marks the trained row nearest its target; the copies row does not touch the floor, so its penetration is not bolded.
Pres.\ is the relative change of cadence and step length from the own-body reference.}
\scriptsize
\setlength{\tabcolsep}{2.5pt}
\begin{tabular}{l|cc|cc|ccc}
\toprule
& \multicolumn{2}{c|}{Lean slope} & \multicolumn{2}{c|}{Preserved slope} & \multicolumn{3}{c}{Other bodies}\\
& fwd & side & cadence & step & Pen.\,\%$\downarrow$ & Pres.$\downarrow$ & Rec.$\uparrow$\\
target & 150 & 65 & 0 & 0 & 0 & 0 & \\
\midrule
\name-X (data) & 150 & 65 & $-0.3$ & $+0.1$ & 0.3 & -- & -- \\
Codec reconstruction & 155 & 55 & $0.0$ & $0.0$ & 0.5 & -- & --\\
\midrule
Captured grid & 22 & $-8$ & $-0.2$ & $+0.1$ & 60.9 & 0.25 & 0.235\\
Geometric copies & $-77$ & 6 & $+0.7$ & $-0.2$ & 0.0 & 0.97 & 0.341\\
No lean targets & 65 & 12 & $+0.2$ & $-0.1$ & \textbf{0.0} & 0.22 & 0.370\\
\name-X & \textbf{105} & \textbf{52} & $\mathbf{0.0}$ & $\mathbf{0.0}$ & \textbf{0.0} & \textbf{0.12} & \textbf{0.514}\\
\bottomrule
\end{tabular}
\label{tab:data}
\end{table}

The captured grid binds every persona to one body; the cross design breaks that binding in three steps: it copies every clip to every body, the optimizer corrects the copies to the target's geometry and contacts, and the two assigned lean targets add what the correction does not.
Table~\ref{tab:data} trains the same controller after each step and reads the columns that separate them: the four slopes of the body sweep, penetration and the preservation error on the sweep bodies, and persona recovery on the held-out bodies.
Trained on the grid, the controller keeps a seventh of the forward lean, 22$^\circ$ per metre against 150, none of the side lean, and penetrates the floor in 61\% of frames on the sweep bodies and 73\% on the held-out bodies: having seen each performer on one body, it cannot place the feet of another.
Trained on the copies, it barely touches the floor, a contact ratio of 0.21 with its feet 4.5\,cm up, so its zero penetration is no merit, and its lean association runs the wrong way, $-77^\circ$ per metre, because the copies carry none.
Trained on \name-X, it keeps the lean, 105 and 52$^\circ$ per metre, keeps cadence and step length flat, and does not penetrate; its preservation error, 0.12, is the lowest of the three, against 0.25 for the grid and 0.97 for the copies, whose cadence and step drift far from the performer's own.
Trained on the optimizer's output without the two lean targets, it does not penetrate, but its lean slopes are 65 and 12$^\circ$ per metre and its preservation error is 0.22: the geometric correction alone turns the association the right way, and the two targets add 40$^\circ$ per metre to each slope and lower the preservation error to 0.12.
Persona recovery on the held-out bodies rises with each step, from 0.24 to 0.34, 0.37, and 0.51, so the targets carry most of the last gain, and practitioners rate the fit of motion to body higher with them, 4.56 against 3.93 (Section~\ref{sec:eval_user}).

\subsection{Why two stages}
\label{sec:eval_capacity}

We hypothesized that a one-stage pose-space network averages the small effects that persona, body, and style carry, and that the two-stage allocation keeps them.
Two measurements test this: the spread columns of Table~\ref{tab:main} and our backbone trained in pose space.

\begin{figure}[t]
  \centering
  \includegraphics[width=\columnwidth]{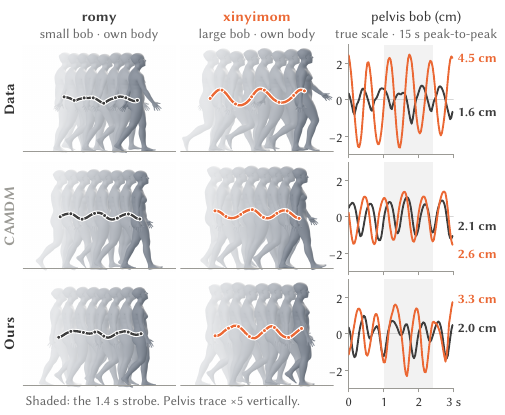}
  \caption{The baseline averages two performers' pelvis bob; ours keeps part of the difference.
  Left: side-view strobes over 1.4\,s, each performer on her own body at a common scale, with the pelvis trace exaggerated five times.
  Right: pelvis height minus its trend at true scale; labels give the peak-to-peak amplitude over 15\,s.}
  \label{fig:averaging}
\end{figure}

\paragraph{Averaging}
Per performer, the generated pelvis bob and jerk can be compared against the same descriptor in that performer's reference, on the own body under one command and style, for CAMDM and for ours.
A controller that keeps the performers puts its points on the diagonal; one that averages them keeps the mean and loses the range.
The same backbone in pose space falls between the two, with 0.61 of the amplitude spread and a style accuracy of 0.37.
Recovery shows it most directly: the baseline's 0.32 on the own body equals that of our prior without a persona input (Table~\ref{tab:ablation_prior}), whereas ours reaches 0.63.
Figure~\ref{fig:averaging} shows it on two performers whose pelvis bob differs by 3.0\,cm in the data: the baseline brings them to within 0.55\,cm, ours keeps 1.3\,cm, under half of the gap.

\paragraph{The same backbone in pose space}
Table~\ref{tab:backbone} trains one transformer on \name-X in three ways: as flow matching in pose space with the same losses and steps as ours, as four-step diffusion in pose space with CAMDM's sampler, and as two-step flow matching in the codec's token space, which is the controller.
Under our training setting and step budget the first does not walk: its feet never register a contact and its root drifts 26\,cm from the command, because its one-step prediction has no contact structure and the error in root height compounds from block to block.
The second walks, likely because its stochastic steps commit to one mode, which is why the baseline is a pose-space diffusion model and not our model without the codec.
It sits closer to the data in pose FPD, but it penetrates the floor in 19\% of frames against 1\%, skates more, 0.57 against 0.49, keeps less of the amplitude spread, 0.61 against 0.70, and reaches a lower style accuracy, 0.37 against 0.74.
The two cost the same 10\,ms per block on an RTX 5080; on two CPU threads the four pose-space passes cost 87\,ms where the token space costs 27.
The middle row is our transformer with CAMDM's sampler, not CAMDM itself, and the comparison holds for this capacity and step budget rather than for pose-space flow matching as such.
\begin{table}[t]
\centering
\caption{One transformer trained three ways on \name-X, own-body set: flow matching in pose space, four-step diffusion in pose space with CAMDM's sampler, and two-step flow matching in the codec's token space, the controller.
Spread is the amplitude group of Table~\ref{tab:main}; ms per block on an RTX 5080 and on two CPU threads.
Bold marks the best of the two rows that walk; the first does not walk under either commit schedule and is reported under the three-frame one.}
\scriptsize
\setlength{\tabcolsep}{1.4pt}
\begin{tabular}{l|cc|ccc|cc|cc}
\toprule
& \multicolumn{2}{c|}{FPD$\downarrow$} & & & & & & \multicolumn{2}{c}{ms}\\
Backbone & pose & contact & Pen.\,\%$\downarrow$ & Skate$\downarrow$ & IoU$\uparrow$ & Style$\uparrow$ & Spread$\uparrow$ & GPU & CPU\\
\midrule
Data & -- & 0.83 & 0.5 & 0.34 & 0.57 & 0.80 & 1.00 & -- & --\\
FM pose, 2 steps & 1.74 & 288 & 2.2 & -- & 0.00 & 0.52 & -- & 6.2 & 44\\
DDPM pose, 4 steps & \textbf{0.28} & 1.90 & 18.9 & 0.57 & 0.49 & 0.37 & 0.61 & 9.9 & 87\\
FM token, 2 steps & 0.32 & \textbf{1.70} & \textbf{1.0} & \textbf{0.49} & \textbf{0.54} & \textbf{0.74} & \textbf{0.70} & 9.9 & \textbf{27}\\
\bottomrule
\end{tabular}
\label{tab:backbone}
\end{table}

\subsection{User study}
\label{sec:eval_user}

The metrics above measure motion; a user study asks whether practitioners see and feel the same.
Twenty-eight professional animators and researchers each drove the two controllers for three minutes, ours and CAMDM side by side in randomly assigned positions, switching persona, body, and style at will.
With the captured motion of the selected performer as the reference, they rated each controller separately on a five-point scale for six criteria: the realism of the motion, its match to the persona and to the style, the artifacts of the body shape, the responsiveness, and how well it follows the command.

Table~\ref{tab:user} reports the means.
Ours is rated higher on five criteria and level on the sixth, the match to the style, 3.91 against 3.82, where both controllers read the same label embedding.
The largest gap is the match to the persona, 4.17 against 2.59, the axis on which the baseline averages, and shape artifacts fall from 2.17 to 1.26; realism, responsiveness, and command following differ by 0.33 to 0.75.
The same participants, in the same blind side-by-side setting, also compared our controller with the same controller trained without the two lean targets (Table~\ref{tab:data}) on every body, the extremes included, and rated how well the motion fits the body: 4.56 with the targets against 3.93 without them.

\begin{table}[t]
\centering
\caption{Practitioners rate ours higher on five criteria and level on style: mean of all ratings collected from 28 professionals on a five-point scale. Lower is better for artifacts; bold marks the better controller.}
\scriptsize
\setlength{\tabcolsep}{4pt}
\begin{tabular}{l|cc}
\toprule
Criterion & Ours & CAMDM\\
\midrule
Realism of the motion $\uparrow$ & \textbf{3.97} & 3.22\\
Match to the persona $\uparrow$ & \textbf{4.17} & 2.59\\
Match to the style $\uparrow$ & \textbf{3.91} & 3.82\\
Body-shape artifacts $\downarrow$ & \textbf{1.26} & 2.17\\
Responsiveness $\uparrow$ & \textbf{3.71} & 3.31\\
Follows the command $\uparrow$ & \textbf{4.12} & 3.79\\
\bottomrule
\end{tabular}
\label{tab:user}
\end{table}

\subsection{Design ablations}
\label{sec:eval_ablation}

Three decisions carry the claims of the design: where the target shape enters, whether the decoder reads the boundary frames, and how the persona is represented.
Tables~\ref{tab:ablation_codec} and~\ref{tab:ablation_prior} retrain one stage with one decision changed; a codec variant gets a prior retrained on it, and the codec table separates frame-aligned reconstruction from rollout, because a decision can leave the first unchanged and alter the second.

\begin{table}[h]
\centering
\caption{Ablations on the codec, on \name-X: one decision changed and the prior retrained.
Reconstruction columns are frame-aligned with the source, with the reference-label IoU; rollout columns follow the protocol on the own-body set.
Lean is the forward-lean slope as a fraction of the data's, after reconstruction and after rollout, with a target of 1; bold marks the best measured row, and for Lean the row nearest 1.
}
\scriptsize
\setlength{\tabcolsep}{2.5pt}
\begin{tabular}{l|cc|ccccc}
\toprule
& \multicolumn{2}{c|}{Reconstruction} & \multicolumn{5}{c}{Rollout}\\
Codec & IoU$\uparrow$ & Lean$\to$1 & Skate$\downarrow$ & IoU$\uparrow$ & Seam$\downarrow$ & Pen.\,\%$\downarrow$ & Lean$\to$1\\
\midrule
Reference & 1.00 & 1.03 & \textbf{0.49} & \textbf{0.54} & \textbf{3.45} & 1.0 & 0.70\\
No $\shapevec$ in the decoder & 1.00 & \textbf{0.99} & 0.55 & 0.35 & 3.46 & 42.0 & 0.61\\
No boundary frames & 1.00 & 1.02 & 0.62 & 0.53 & 3.81 & \textbf{0.9} & \textbf{1.17}\\
\bottomrule
\end{tabular}
\label{tab:ablation_codec}
\end{table}

\paragraph{Shape conditioning}
The target shape does a different job in each stage.
In the decoder it places the feet: without it the controller penetrates the floor in 42\% of frames against 1\%, and its contact consistency falls from 0.54 to 0.35, although reconstruction is unchanged.
In the prior it plans path and feet: without the shape token the skate rises from 0.49 to 0.65 and the root error from 5.5 to 7.8\,cm, while the spread between performers grows, since a prior blind to the body varies more freely and lands less.
A linear probe from the tokens to $\shapevec$ gives an $R^2$ of $-0.39$, yet decoding a clip's tokens on another body recovers 98\% of the gap between its two retargeted versions: the tokens are reusable across bodies but not shape-free.
\paragraph{Which stage carries the lean}
When only one stage receives the target body, under a command fixed in metres rather than in body units, the decoder alone reproduces 78 of the 125$^\circ$ per metre that both stages produce and the prior alone reverses it, while side lean is shared; the tokens have to be generated for the body they are rendered on, so both stages keep the shape.
\paragraph{Boundary frames}
The boundary frames make the blocks join: without them the seam grows from 3.45 to 3.81\,cm per frame and the skate from 0.49 to 0.62, while reconstruction is unchanged; with them the boundary jump is that of an ordinary frame.
\paragraph{Persona conditioning}
The prior represents a persona by one ID token and three typed attribute tokens, with style separate.
Removing the persona collapses the spread between performers from 0.70 to 0.19, lowers the style accuracy from 0.74 to 0.53, and halves the own-body recovery, from 0.63 to 0.32, so both columns measure the persona input and not the body.
The ID alone keeps recovery and spread, 0.64 and 0.69, but loses style accuracy, 0.66; the attributes alone lose all three, 0.51, 0.62, and 0.56, bounded by the eight cells that hold more than one performer; and a CLIP embedding of a persona prompt keeps spread and style accuracy, 0.70 and 0.73, but loses recovery, 0.52.

\begin{table}[t]
\centering
\caption{Ablations on the prior, on \name-X, own-body set: one decision changed per row on the final codec.
Track.\ is the root error in cm, Rec.\ persona recovery, Style the style accuracy, Spread the amplitude group, Lean the forward-lean slope as a fraction of the data's on the sweep bodies, with a target of 1; bold marks the best measured row, and for Lean the row nearest 1.
Rows within 0.03 of skate or 0.04 of recovery of the reference are within one training seed's spread and are not ranked.}
\scriptsize
\setlength{\tabcolsep}{1.05pt}
\begin{tabular}{l|cc|ccc|ccc|c}
\toprule
& \multicolumn{2}{c|}{FPD$\downarrow$} & & & & & & & \\
Prior & pose & contact & Skate$\downarrow$ & IoU$\uparrow$ & Track.$\downarrow$ & Rec.$\uparrow$ & Style$\uparrow$ & Spread$\uparrow$ & Lean$\to$1\\
\midrule
Reference & 0.32 & 1.70 & 0.49 & \textbf{0.54} & 5.5 & 0.632 & 0.74 & 0.70 & 0.70\\
Dropout 0 & 0.29 & 2.23 & 0.53 & 0.48 & \textbf{2.9} & 0.494 & 0.73 & 0.66 & 1.28\\
No shape token & 0.33 & 1.90 & 0.65 & 0.48 & 7.8 & 0.629 & 0.73 & \textbf{0.76} & 0.61\\
No persona & 0.48 & 2.24 & 0.52 & 0.51 & 5.9 & 0.318 & 0.53 & 0.19 & \textbf{1.00}\\
ID only & 0.32 & 1.70 & 0.49 & \textbf{0.54} & 5.4 & 0.644 & 0.66 & 0.69 & 0.82\\
Attributes only & 0.39 & 2.00 & 0.50 & 0.53 & 5.7 & 0.507 & 0.56 & 0.62 & 0.74\\
CLIP prompt persona & 0.34 & 1.74 & 0.52 & 0.53 & 6.0 & 0.516 & 0.73 & 0.70 & 0.78\\
4 blocks $\times$ 256 & 0.35 & 1.87 & 0.50 & 0.48 & 4.2 & 0.613 & 0.72 & 0.61 & 0.81\\
Time token, no DiT blocks & 0.35 & 1.73 & 0.46 & \textbf{0.54} & 3.7 & 0.669 & 0.75 & 0.72 & 0.80\\
History token never dropped & 0.32 & 1.60 & 0.47 & \textbf{0.54} & 5.4 & 0.640 & \textbf{0.76} & 0.69 & 0.82\\
Mirror motion only & 0.30 & 1.63 & 0.48 & \textbf{0.54} & 4.7 & \textbf{0.676} & \textbf{0.76} & 0.71 & 0.78\\
No mirroring & \textbf{0.26} & \textbf{1.54} & 0.48 & 0.52 & 3.3 & 0.642 & \textbf{0.76} & 0.70 & 0.90\\
\bottomrule
\end{tabular}
\label{tab:ablation_prior}
\end{table}

\section{Applications}
\label{sec:applications}

The two stages take their conditions as tokens, so a new task adds a token and retrains the prior; the codec, the cross design, and the single-axis test stay as they are.
We outline two such extensions as preliminary directions, not as validated contributions, each with one worked example.

\begin{figure}[t]
  \centering
  \includegraphics[width=\columnwidth]{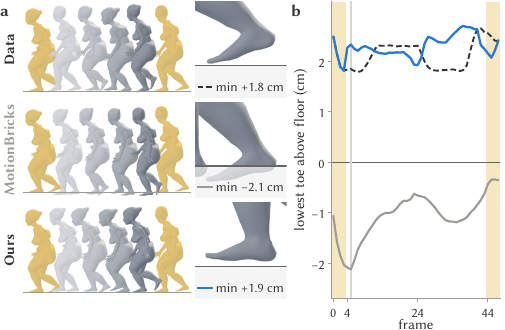}
  \caption{In-betweening on a 105\,cm body: MotionBricks~\cite{wang2026motionbricks}, which has no body input, sinks the feet into the floor for the whole gap, whereas ours keeps them where the data has them in this example.
  Both receive the same keyframes (gold) and root path of a backward walk and generate the 40 frames between them.
  (a) Poses spread horizontally for legibility, with the feet shown at frame 5.
  (b) Lowest toe joint height: minimum $+1.8$\,cm in the data, $+1.9$\,cm for ours, and $-2.1$\,cm for MotionBricks.}
  \label{fig:inbetween}
\end{figure}

\paragraph{Character-aware in-betweening}
Given the pose at the start of a gap and a keyframe at its end, the prior fills the gap for a chosen persona on a chosen body: the frozen codec encodes the end keyframe into one more token, built like the history token, and everything else is the controller of Section~\ref{sec:method}.
Because every clip of \name-X exists on every body, one pair of keyframes has a frame-aligned answer on a child and on a tall adult, so the filled motion can be scored on bodies the performer never had (Figure~\ref{fig:inbetween}).

\begin{figure}[t]
  \centering
  \includegraphics[width=\columnwidth]{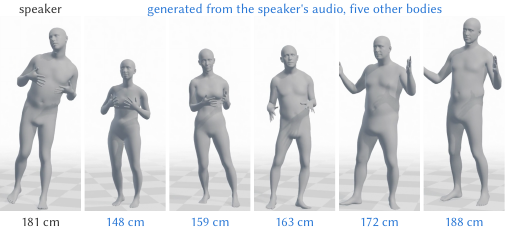}
  \caption{Speech-driven gesture: one speech clip of a BEAT2 speaker drives five other speakers on their own bodies, short to tall, each in its own manner.
  Left, the speaker's motion capture; feet locked after generation.}
  \label{fig:gesture}
\end{figure}

\paragraph{Speech-driven gesture}
The same three axes exist in BEAT~\cite{liu2022beat}: thirty speakers with distinct bodies, eight emotions, and speech aligned to the motion, with SMPL-X bodies in its second release~\cite{liu2024emage}.
Persona becomes the speaker, body the speaker's shape, style the emotion, and an audio token replaces the trajectory; Figure~\ref{fig:gesture} drives five speakers with the speech of a sixth.

\section{Discussion and Limitations}
\label{sec:discussion}

Capture alone could not separate who is moving from the body that moves, because every performer came with one body, and the cross design supplies the missing counterfactual by construction: it is a specification, and what the cross-body experiments verify is that the controller follows it, not how a performer would in fact move in another body.
The single-axis test shows that a controller trained on it satisfies the split in part: the body moves the trunk lean in the direction of the two associations the grid revealed, by three quarters and two fifths of their slopes, and leaves cadence and normalized step length alone, and persona and style each move the descriptors the grid assigns to them while neither moves what the command fixes; what the controller keeps of the data's spread is uneven, most of step length, pelvis bob, and trunk lean between performers, little of jerk and arm swing, and little of pelvis bob between styles.
The effects that carry identity are a few degrees and a few centimetres, and a one-stage pose-space controller flattens part of them across 44 performers and loses the floor away from mid-height bodies; a shape-aware codec for geometry and contact and a prior for selection in its token space keep more of them and the floor, at two steps and 27\,ms per block on two CPU threads.
One controller then covers a captured catalog that would otherwise need per-character asset sets, and the measurement behind it, one repeated grid and one single-axis test, is not specific to locomotion.

Three limits bound what the controller can do.

\paragraph{A new persona needs an example}
A persona is one of the 44 captured performers, selected by ID together with its three attributes.
The attributes explain little of a performer's signature, so a new combination of attributes does not define a new persona, and the controller has no text or parameter interface for one.
Adding a persona means obtaining example motion of that person and training their ID in; the cross design then puts them on every body.
Example motion is the right interface in our view, because the signature lives in timing, contact, and posture that a description does not pin down; an encoder that reads it from a short clip instead of a trained ID is the natural next step.

\paragraph{Physically informed, not simulated}
The retargeter keeps bodies out of each other and out of the floor, keeps a quasi-static balance proxy inside the support, and budgets effort against the source, and the controller inherits these properties from its data.
None of this is a simulation: there are no forces or torques, balance is checked kinematically, and the effort budget is a proxy.
The adjustments the optimizer makes when a motion is carried to another body are therefore correct in geometry and plausible in balance, but not established as dynamically correct; a heavy body's run can be feasible in our terms and not in a simulator's.

\paragraph{Scale and coverage}
The annotated grid holds 44 performers, nine styles, and seven commands, all locomotion, about 33 hours in all; \name-X multiplies the bodies, not the performers or the actions.
The persona catalog therefore ends at 44, the styles at nine, and the catalog covers no gesture, interaction, or action beyond locomotion; the gesture extension of Section~\ref{sec:applications} would borrow other data for that.
Growing any of these axes needs new repeated capture of the same performers under the same conditions; the cross design and the single-axis test carry over unchanged.

\section{Conclusion}
We asked how a locomotion controller can preserve who is moving while changing the body that carries the motion.
One-body-per-performer capture observes only the diagonal of the persona--body grid, so we wrote down a conservative, auditable specification from a repeated grid of 44 performers: two trunk-lean associations assigned to shape, and preservation targets for quantities without a detectable shape association.
A physically informed retargeter instantiated it as a persona--body cross design, on which we trained a shape-aware codec and a co-conditioned latent flow-matching prior, one real-time controller for the captured catalog.
Single-axis interventions show that the controller follows the selected targets in part, keeping most of the forward-lean association and part of the side-lean one.

\bibliographystyle{ACM-Reference-Format}
\bibliography{main.bib}

\clearpage
\appendix

\section{Dataset details}
\label{sec:sup_dataset}

\begin{table*}[t]
    \centering
    \caption{Example annotations.
    Each clip is linked to a closed-set performer ID, three typed persona attributes, the performer's source-body metadata, a style label, and a locomotion command.
    Beta is the 10-dimensional SMPL-X shape vector; age and gender describe the capture cohort but are not persona conditions.}
    \resizebox{\textwidth}{!}{
    \begin{tabular}{l|cccccccc|cc|c}
        \toprule
        & \multicolumn{8}{c|}{Performer profile and source body} & \multicolumn{2}{c|}{Style} & Command \\ \midrule
        Take name & Age & Gender & Height (cm) & Weight (kg) & Beta & Role & Affiliation & Dominance & Label & Group & Base movement \\
        \midrule
        s001\_angry\_fw     & 5 & M & 112 & 17 & [-5.68, 1.26, ...] & Child & Reserved & Compliant & Angry & Affective & forward walking \\
        s015\_neutral\_bw     & 22 & F & 165 & 47 & [0.96, 0.27, ...] & Student & Sociable & Assertive & Neutral & Affective & backward walking \\
        s032\_twofootjump\_fr & 23 & M & 189 & 80 & [1.44, -1.35, ...] & Athlete & Gregarious & Dominant & Two-foot jump & Locomotion variant & forward running \\
        \bottomrule
    \end{tabular}
    }
    \label{tab:annotation_example}
\end{table*}

\begin{figure}[t]
	\centering
	\includegraphics[width=1.\columnwidth]{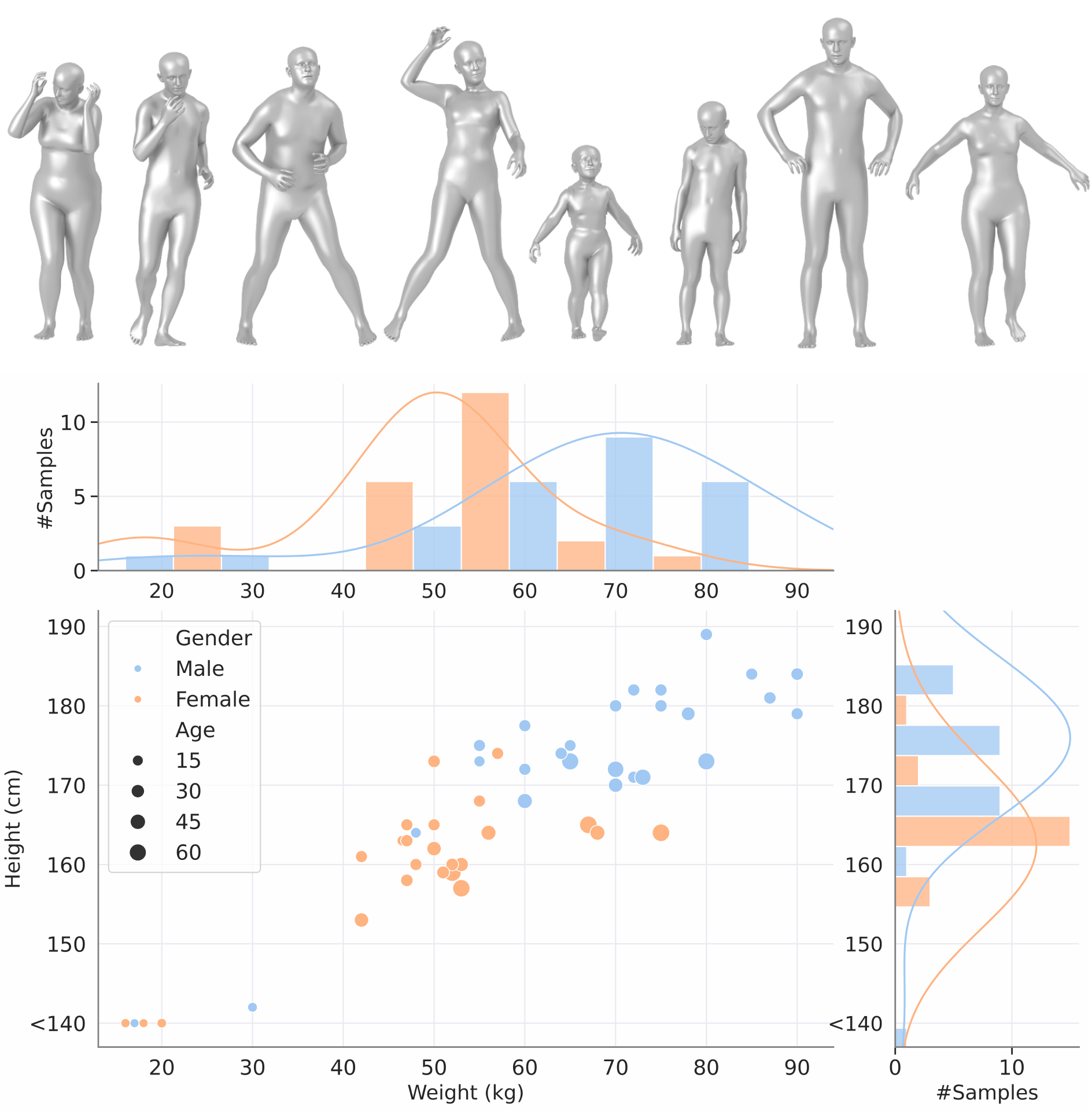}
	\caption{Top: eight of the SMPL-X bodies fitted to the performers.
    Bottom: the distribution of height, weight, and age.
    Blue points are male performers, orange points female.}
	\label{fig:sup_dataset}
\end{figure}

Table~\ref{tab:sup_dataset_comp} places \name next to the locomotion datasets that controllers are usually trained on.
Figure~\ref{fig:sup_dataset} shows eight of the fitted SMPL-X bodies and the distribution of height, weight, and age over the performers.

\begin{table*}[t]
\caption{Comparison with locomotion datasets.
\name combines performer diversity, repeated styles, repeated commands, and SMPL-X bodies.
``Text'' denotes performer- or style-level annotation rather than motion-content captions.}
\resizebox{\textwidth}{!}{
\begin{tabular}{l|c|cccc|c|cccc|ccc}
\toprule
& & \multicolumn{4}{c|}{Locomotion coverage} & & \multicolumn{4}{c|}{Performers} & \multicolumn{3}{c}{Scale and annotation}\\
Dataset & Accessible & Forward & Backward & Sideways & Fingers & \#Perf. & Age & Height & Weight & Text & \#Seq & Hours & SMPL\\
\midrule
Edinburgh~\shortcite{pfnn} & \cmark & \cmark & \cmark & \cmark & \xmark & -- & \xmark & \xmark & \xmark & \xmark & 80 & 1 & \xmark\\
LAFAN1~\shortcite{harveyRobustMotionInbetweening2020} & \cmark & \cmark & \cmark & \cmark & \xmark & 5 & \xmark & \xmark & \xmark & \xmark & 77 & 4.6 & \xmark\\
BFA~\shortcite{abermanUnpairedMotionStyle2020} & \cmark & \cmark & \xmark & \xmark & \xmark & 1 & \xmark & \xmark & \xmark & \xmark & 33 & 1.5 & \xmark\\
100STYLE~\shortcite{100style} & \cmark & \cmark & \cmark & \cmark & \xmark & 1 & \xmark & \xmark & \xmark & \xmark & 810 & 18.75 & \xmark\\
Multi-sub~\shortcite{hou2024causal} & \xmark & \cmark & \cmark & \xmark & \xmark & 12 & \xmark & 154--195 & \xmark & \xmark & -- & 4 & \xmark\\
PerMo~\shortcite{kim2025personabooth} & \cmark & \cmark & \xmark & \xmark & \xmark & 5 & \xmark & \xmark & \xmark & \cmark & 6610 & 8.5 & \cmark\\
\name & -- & \cmark & \cmark & \cmark & \cmark & 44 (+4 bodies) & 5--68 & 105--189 & 16--90 & \cmark & 2,590 & 33 & \cmark\\
\bottomrule
\end{tabular}}
\label{tab:sup_dataset_comp}
\end{table*}

\section{The measurement in full}
\label{sec:sup_measurement}

Section~4.4 of the main paper reports the measured body effects.
This section gives the analysis behind them, so that every number in the paper can be traced to a table here.

\subsection{Design and descriptors}
The grid, as analysed, holds the 2,461 clips available for the 63 style $\times$ command cells.
Table~\ref{tab:sup_clips} lists the clips of every performer in every style.

\begin{table*}[t]
\centering
\scriptsize
\caption{Clips per performer and style in the analysed grid; each style holds at most seven clips, one per command.
Sixteen performers fill all 63 cells, and P33, P43, P37, P03, and P29 miss many.
Columns are angry, big step, depressed, drunk, fear, happy, neutral, swimming, and two-foot jump.}
\setlength{\tabcolsep}{3.2pt}
\begin{tabular}{l*{9}{c}c@{\hspace{12pt}}l*{9}{c}c}
\toprule
Performer & Ang & Big & Dep & Dru & Fea & Hap & Neu & Swi & Jmp & Sum & Performer & Ang & Big & Dep & Dru & Fea & Hap & Neu & Swi & Jmp & Sum\\
\midrule
P01 & 6 & 6 & 6 & 7 & 6 & 6 & 6 & 6 & 6 & 55 & P23 & 6 & 6 & 6 & 6 & 6 & 6 & 6 & 6 & 0 & 48\\
P02 & 7 & 7 & 7 & 7 & 7 & 7 & 7 & 7 & 7 & 63 & P24 & 7 & 7 & 7 & 7 & 7 & 7 & 7 & 7 & 7 & 63\\
P03 & 6 & 4 & 5 & 3 & 3 & 1 & 7 & 1 & 5 & 35 & P25 & 7 & 7 & 7 & 7 & 7 & 7 & 7 & 7 & 7 & 63\\
P04 & 7 & 7 & 7 & 7 & 7 & 7 & 6 & 7 & 7 & 62 & P26 & 7 & 7 & 7 & 7 & 7 & 7 & 6 & 7 & 7 & 62\\
P05 & 7 & 7 & 7 & 7 & 7 & 7 & 7 & 7 & 7 & 63 & P27 & 7 & 7 & 7 & 7 & 7 & 7 & 7 & 7 & 7 & 63\\
P06 & 7 & 7 & 7 & 7 & 7 & 7 & 7 & 7 & 7 & 63 & P28 & 7 & 7 & 7 & 7 & 7 & 7 & 7 & 7 & 7 & 63\\
P07 & 7 & 7 & 7 & 7 & 7 & 7 & 7 & 7 & 6 & 62 & P29 & 7 & 4 & 7 & 7 & 1 & 7 & 7 & 0 & 3 & 43\\
P08 & 7 & 7 & 7 & 7 & 6 & 7 & 7 & 7 & 7 & 62 & P30 & 7 & 7 & 7 & 6 & 7 & 7 & 7 & 7 & 2 & 57\\
P09 & 6 & 6 & 6 & 6 & 5 & 6 & 6 & 6 & 6 & 53 & P31 & 7 & 7 & 7 & 7 & 7 & 7 & 7 & 7 & 6 & 62\\
P10 & 7 & 7 & 7 & 7 & 6 & 7 & 7 & 7 & 7 & 62 & P32 & 6 & 6 & 6 & 6 & 6 & 5 & 6 & 6 & 6 & 53\\
P11 & 7 & 7 & 7 & 7 & 7 & 7 & 7 & 7 & 7 & 63 & P33 & 0 & 0 & 0 & 0 & 0 & 0 & 0 & 0 & 1 & 1\\
P12 & 7 & 7 & 7 & 7 & 7 & 7 & 7 & 7 & 7 & 63 & P34 & 7 & 7 & 7 & 7 & 7 & 7 & 7 & 6 & 7 & 62\\
P13 & 6 & 6 & 6 & 6 & 6 & 6 & 6 & 7 & 6 & 55 & P35 & 7 & 7 & 7 & 7 & 7 & 7 & 7 & 7 & 7 & 63\\
P14 & 7 & 7 & 7 & 7 & 7 & 7 & 7 & 7 & 7 & 63 & P36 & 7 & 7 & 7 & 7 & 7 & 7 & 6 & 7 & 6 & 61\\
P15 & 7 & 7 & 7 & 7 & 7 & 7 & 6 & 7 & 7 & 62 & P37 & 3 & 1 & 3 & 0 & 0 & 4 & 6 & 3 & 1 & 21\\
P16 & 7 & 7 & 7 & 7 & 7 & 7 & 7 & 6 & 7 & 62 & P38 & 7 & 7 & 7 & 6 & 7 & 7 & 7 & 7 & 7 & 62\\
P17 & 7 & 7 & 7 & 7 & 7 & 7 & 7 & 7 & 7 & 63 & P39 & 7 & 7 & 7 & 7 & 7 & 6 & 7 & 7 & 7 & 62\\
P18 & 6 & 6 & 6 & 6 & 6 & 6 & 6 & 6 & 6 & 54 & P40 & 7 & 7 & 7 & 7 & 7 & 7 & 7 & 7 & 7 & 63\\
P19 & 7 & 7 & 7 & 7 & 7 & 7 & 7 & 7 & 7 & 63 & P41 & 7 & 7 & 7 & 7 & 7 & 7 & 7 & 7 & 7 & 63\\
P20 & 7 & 7 & 7 & 7 & 6 & 7 & 7 & 6 & 7 & 61 & P42 & 6 & 6 & 6 & 6 & 5 & 6 & 6 & 6 & 6 & 53\\
P21 & 7 & 7 & 7 & 7 & 7 & 7 & 7 & 7 & 7 & 63 & P43 & 0 & 0 & 0 & 0 & 0 & 0 & 2 & 0 & 0 & 2\\
P22 & 7 & 6 & 7 & 7 & 7 & 7 & 7 & 7 & 2 & 57 & P44 & 6 & 7 & 7 & 7 & 7 & 7 & 7 & 7 & 7 & 62\\
\cmidrule(l){12-22}
\multicolumn{11}{l}{} & All 44 & 281 & 274 & 281 & 275 & 265 & 276 & 284 & 268 & 257 & 2,461\\
\bottomrule
\end{tabular}
\label{tab:sup_clips}
\end{table*}

The 27 descriptors are computed in the pelvis-facing frame from forward-kinematics joint positions.
Twenty-three describe the motion and four are reference quantities used only as checks.
They cover timing (cadence, contact ratio, double support, step regularity), space (speed, step length, step width, foot clearance), the root and the limbs (pelvis bob and sway; knee range, stance knee, hip range; arm swing, elbow angle, hand distance, hand reach), and the trunk, the head, and energy (forward and side lean, head sway and bob, joint speed, jerk).
Lengths are normalized by the performer's own leg length, hip width, or shoulder width, and cadence is taken from the autocorrelation of ankle height.
Every clip is read from its BVH file with the static first frame dropped, clips shorter than 90 frames are excluded, and the descriptors are also computed on non-overlapping 10-second windows to estimate the within-clip noise floor.
Table~\ref{tab:sup_descriptors} lists the definition, estimator, and unit of each descriptor.

\begin{table*}[t]
\centering
\scriptsize
\caption{The 27 gait descriptors, computed in the pelvis-facing frame from forward-kinematics joint positions.
Lengths are divided by the performer's leg length, hip width, or half shoulder width, so similarity scaling leaves them unchanged.
Daggers mark the four reference quantities, which enter only the variance decomposition and the first body fits, as checks.}
\begin{tabular}{l l p{10cm} l}
\toprule
Group & Descriptor & Definition and estimator & Unit\\
\midrule
Timing & Cadence & Steps per second, $2/T$, where the stride period $T$ is the first autocorrelation peak of each ankle's height; the median over 10-second sliding windows & steps/s\\
& Cadence (events)$^\dagger$ & Number of contact onsets of both feet divided by the duration; kept only as a check, since it undercounts steps & steps/s\\
& Contact ratio & Fraction of frames in which a foot is in contact; contact starts when the ankle or toe is lower than $\max(3\%\ \text{leg}, 2\,\text{cm})$ and ends when both are higher than $\max(6\%\ \text{leg}, 4\,\text{cm})$ & fraction\\
& Contact ratio (labels)$^\dagger$ & The same from the stored contact labels (foot speed below a threshold and height at most 3\,cm), which miss most stance frames in running & fraction\\
& Double support & Fraction of frames with both feet in contact & fraction\\
& Step regularity & Coefficient of variation of the step intervals (higher is less regular) & --\\
\midrule
Space & Speed & Root speed divided by leg length & leg/s\\
& Step length & Distance covered per step divided by leg length & leg\\
& Step width & Lateral distance between the ankles at contact onset divided by hip width & hip width\\
& Foot clearance & 95th percentile of swing-phase ankle height minus 5th percentile of stance-phase ankle height, divided by leg length & leg\\
\midrule
Root & Pelvis bob & Standard deviation of the high-passed root height, divided by leg length & leg\\
& Pelvis sway & Standard deviation of the high-passed lateral root position, divided by leg length & leg\\
& Pelvis height$^\dagger$ & Root height divided by leg length, a skeletal proportion rather than a motion & leg\\
& Yaw rate$^\dagger$ & Yaw angular speed of the root, which reflects the capture path & deg/s\\
\midrule
Legs & Knee range & 95th minus 5th percentile of the knee angle & deg\\
& Stance knee & Knee flexion during stance & deg\\
& Hip range & Range of the thigh's sagittal angle & deg\\
\midrule
Arms & Arm swing & Range of the shoulder-to-wrist sagittal angle & deg\\
& Elbow angle & Mean elbow flexion & deg\\
& Hand distance & Lateral distance of the wrist from the midline divided by half the shoulder width & half shoulder\\
& Hand reach & Forward position of the wrist divided by leg length & leg\\
\midrule
Trunk and head & Forward lean & Forward tilt of the pelvis-to-neck segment & deg\\
& Side lean & Absolute sideways tilt of the pelvis-to-neck segment & deg\\
& Head sway & Standard deviation of the lateral head position, divided by leg length & leg\\
& Head bob & Standard deviation of the high-passed head height, divided by leg length & leg\\
\midrule
Energy & Joint speed & Mean root-relative joint speed in the facing frame, divided by leg length & leg/s\\
& Jerk & Mean third difference of the joint positions, divided by leg length & leg/s$^3$\\
\bottomrule
\end{tabular}
\label{tab:sup_descriptors}
\end{table*}

\subsection{Variance decomposition}
\begin{table}[t]
\centering
\caption{Variance decomposition on the grid, as fractions of total variance under the additive model in the main paper.
Residual is what the four terms leave; for cadence most of it is estimator variance.
Noise is the robust within-clip floor, from the median absolute deviation over 10-second segments.
Reliability is the split-half correlation of the performer effect, across style groups and across walking versus running.}
\resizebox{\columnwidth}{!}{
\begin{tabular}{l|cccc|c|c|c}
\toprule
Descriptor & Style & Command & Performer & Interaction & Residual & Noise & Reliability\\
\midrule
Speed & 0.02 & 0.21 & \textbf{0.52} & 0.10 & 0.16 & 0.09 & 0.86--0.93\\
Jerk & 0.03 & 0.13 & \textbf{0.51} & 0.20 & 0.14 & 0.02 & 0.76--0.96\\
Sway & 0.02 & 0.21 & \textbf{0.47} & 0.11 & 0.20 & 0.09 & 0.86--0.87\\
Contact ratio & 0.10 & 0.28 & \textbf{0.33} & 0.09 & 0.20 & 0.09 & 0.87--0.92\\
Forward lean & 0.25 & 0.02 & 0.25 & \textbf{0.36} & 0.13 & 0.04 & 0.73--0.94\\
Cadence & 0.04 & 0.16 & 0.21 & 0.18 & 0.42 & 0.02 & 0.73--0.74\\
Arm swing & 0.27 & 0.02 & 0.19 & \textbf{0.32} & 0.20 & 0.07 & 0.62--0.89\\
Pelvis bob & \textbf{0.44} & 0.11 & 0.17 & 0.13 & 0.15 & 0.02 & 0.74--0.84\\
\bottomrule
\end{tabular}}
\label{tab:sup_decomp}
\end{table}

We fit the additive decomposition from the main paper by sequential sums of squares in two orders, and estimate the noise floor from 10-second segments within a clip.
Table~\ref{tab:sup_decomp_full} lists both orders, the raw and robust noise floors, and the split-half reliability for all 27 descriptors.

\begin{table*}[t]
\centering
\scriptsize
\caption{Variance decomposition of all 27 descriptors, as fractions of the total sum of squares.
The performer share exceeds the robust noise floor for every descriptor except step regularity and head sway, and changes by at most 0.01 when the performer term is fitted first.
Interaction and residual are identical in both orders and listed once; the robust noise floor uses a MAD-based variance, because about 3\% of the 10-second cadence and step-length estimates are octave errors.}
\begin{tabular}{l ccccc ccc cc cc}
\toprule
& \multicolumn{5}{c}{Order A: style first} & \multicolumn{3}{c}{Order B: performer first} & \multicolumn{2}{c}{Noise floor} & \multicolumn{2}{c}{Split-half reliability}\\
\cmidrule(lr){2-6}\cmidrule(lr){7-9}\cmidrule(lr){10-11}\cmidrule(l){12-13}
Descriptor & Style & Command & Performer & Interaction & Residual & Performer & Command & Style & Raw & Robust & Styles & Walk/run\\
\midrule
Cadence & 0.04 & 0.16 & 0.21 & 0.18 & 0.42 & 0.21 & 0.16 & 0.04 & 0.65 & 0.02 & 0.73 & 0.74\\
Cadence (events)$^\dagger$ & 0.04 & 0.27 & 0.30 & 0.12 & 0.27 & 0.30 & 0.27 & 0.04 & 0.22 & 0.11 & 0.81 & 0.88\\
Speed & 0.02 & 0.21 & 0.52 & 0.10 & 0.16 & 0.52 & 0.20 & 0.03 & 0.24 & 0.09 & 0.86 & 0.93\\
Step length & 0.16 & 0.15 & 0.27 & 0.12 & 0.30 & 0.27 & 0.14 & 0.17 & 0.54 & 0.12 & 0.90 & 0.87\\
Contact ratio & 0.10 & 0.28 & 0.33 & 0.09 & 0.20 & 0.33 & 0.28 & 0.09 & 0.24 & 0.09 & 0.92 & 0.87\\
Contact ratio (labels)$^\dagger$ & 0.01 & 0.23 & 0.54 & 0.05 & 0.17 & 0.53 & 0.24 & 0.01 & 0.55 & 0.18 & 0.96 & 0.89\\
Double support & 0.11 & 0.30 & 0.28 & 0.10 & 0.21 & 0.28 & 0.30 & 0.11 & 0.24 & 0.11 & 0.90 & 0.87\\
Step regularity & 0.06 & 0.17 & 0.12 & 0.16 & 0.49 & 0.12 & 0.18 & 0.06 & 0.92 & 0.54 & 0.70 & 0.67\\
Step width & 0.04 & 0.59 & 0.09 & 0.07 & 0.22 & 0.09 & 0.59 & 0.04 & 0.34 & 0.07 & 0.85 & 0.84\\
Foot clearance & 0.11 & 0.17 & 0.28 & 0.18 & 0.26 & 0.28 & 0.17 & 0.11 & 0.20 & 0.05 & 0.81 & 0.96\\
Pelvis bob & 0.44 & 0.11 & 0.17 & 0.13 & 0.15 & 0.17 & 0.11 & 0.44 & 0.08 & 0.02 & 0.74 & 0.84\\
Pelvis sway & 0.02 & 0.21 & 0.47 & 0.11 & 0.20 & 0.47 & 0.20 & 0.02 & 0.25 & 0.09 & 0.87 & 0.86\\
Pelvis height$^\dagger$ & 0.07 & 0.01 & 0.78 & 0.08 & 0.06 & 0.78 & 0.01 & 0.07 & 0.05 & 0.02 & 0.98 & 0.98\\
Knee range & 0.05 & 0.25 & 0.30 & 0.15 & 0.24 & 0.30 & 0.26 & 0.05 & 0.20 & 0.07 & 0.85 & 0.86\\
Stance knee & 0.15 & 0.21 & 0.28 & 0.16 & 0.20 & 0.27 & 0.22 & 0.16 & 0.28 & 0.10 & 0.85 & 0.83\\
Hip range & 0.25 & 0.15 & 0.25 & 0.13 & 0.23 & 0.24 & 0.15 & 0.25 & 0.18 & 0.06 & 0.85 & 0.85\\
Arm swing & 0.27 & 0.02 & 0.19 & 0.32 & 0.20 & 0.19 & 0.02 & 0.27 & 0.35 & 0.07 & 0.62 & 0.89\\
Elbow angle & 0.30 & 0.07 & 0.10 & 0.33 & 0.21 & 0.10 & 0.07 & 0.30 & 0.13 & 0.03 & 0.60 & 0.57\\
Hand distance & 0.24 & 0.01 & 0.27 & 0.32 & 0.16 & 0.27 & 0.01 & 0.23 & 0.31 & 0.11 & 0.79 & 0.93\\
Hand reach & 0.36 & 0.03 & 0.16 & 0.32 & 0.14 & 0.16 & 0.03 & 0.36 & 0.15 & 0.06 & 0.60 & 0.89\\
Forward lean & 0.25 & 0.02 & 0.25 & 0.36 & 0.13 & 0.24 & 0.02 & 0.25 & 0.14 & 0.04 & 0.73 & 0.94\\
Side lean & 0.12 & 0.06 & 0.27 & 0.27 & 0.28 & 0.26 & 0.06 & 0.13 & 0.50 & 0.20 & 0.77 & 0.86\\
Head sway & 0.24 & 0.05 & 0.19 & 0.25 & 0.27 & 0.19 & 0.05 & 0.24 & 0.42 & 0.20 & 0.64 & 0.88\\
Head bob & 0.40 & 0.08 & 0.17 & 0.18 & 0.18 & 0.17 & 0.08 & 0.40 & 0.10 & 0.02 & 0.67 & 0.84\\
Joint speed & 0.14 & 0.13 & 0.41 & 0.16 & 0.16 & 0.41 & 0.13 & 0.15 & 0.20 & 0.07 & 0.76 & 0.91\\
Jerk & 0.03 & 0.13 & 0.51 & 0.20 & 0.14 & 0.51 & 0.13 & 0.04 & 0.07 & 0.02 & 0.76 & 0.96\\
Yaw rate$^\dagger$ & 0.05 & 0.18 & 0.43 & 0.12 & 0.21 & 0.44 & 0.17 & 0.06 & 0.37 & 0.23 & 0.85 & 0.93\\
\bottomrule
\end{tabular}
\label{tab:sup_decomp_full}
\end{table*}

\subsection{What the body predicts}
Every performer effect is regressed on the body measures with leave-one-performer-out ridge regression, on all performers and within the adults.
Table~\ref{tab:sup_body_r2} lists these fits, the strongest single adult correlate, and the fits of the three attributes for all 27 descriptors.

\begin{table*}[t]
\centering
\scriptsize
\caption{Body measures and attributes predict little of the performer effect.
Values are leave-one-performer-out $R^2$, negative when worse than the mean; full uses the nine body measures, compact leg length, BMI, child, and female, and nested selects one or two measures inside each training fold (reference quantities not fitted).
Attributes are the role one-hot with ordinal affiliation and dominance; the last column fits them to the residual of the better body model.}
\begin{tabular}{l cc cc cc l ccc}
\toprule
& \multicolumn{2}{c}{Body, all 44} & \multicolumn{2}{c}{Body, 39 adults} & \multicolumn{2}{c}{Body, nested} & & \multicolumn{3}{c}{Attributes}\\
\cmidrule(lr){2-3}\cmidrule(lr){4-5}\cmidrule(lr){6-7}\cmidrule(l){9-11}
Descriptor & Full & Compact & Full & Compact & All & Adults & Strongest adult correlate ($r$) & All & Adults & Body residual\\
\midrule
Cadence & $-$0.03 & $-$0.17 & $-$0.00 & $-$0.08 & $-$0.12 & $-$0.03 & arm/leg ($-$0.20) & $-$0.01 & $-$0.02 & $-$0.12\\
Cadence (events)$^\dagger$ & $-$0.08 & $-$0.11 & $-$0.04 & $-$0.02 & -- & -- & BMI ($-$0.32) & 0.08 & 0.09 & 0.07\\
Speed & $-$0.30 & $-$0.23 & 0.06 & $-$0.04 & $-$0.13 & $-$0.07 & arm/leg ($-$0.25) & 0.02 & 0.07 & $-$0.06\\
Step length & $-$0.29 & $-$0.27 & 0.24 & 0.20 & $-$0.32 & 0.22 & BMI ($-$0.33) & 0.00 & 0.09 & $-$0.11\\
Contact ratio & $-$0.07 & $-$0.07 & $-$0.12 & $-$0.08 & $-$0.16 & 0.07 & arm/leg ($-$0.33) & $-$0.22 & $-$0.23 & $-$0.21\\
Contact ratio (labels)$^\dagger$ & $-$0.18 & $-$0.06 & 0.02 & $-$0.04 & -- & -- & BMI ($+$0.33) & $-$0.18 & $-$0.15 & $-$0.12\\
Double support & $-$0.08 & $-$0.08 & $-$0.11 & $-$0.07 & 0.03 & $-$0.04 & BMI ($+$0.28) & $-$0.17 & $-$0.16 & $-$0.16\\
Step regularity & $-$0.03 & $-$0.05 & $-$0.24 & $-$0.08 & $-$0.14 & $-$0.05 & hip w. ($+$0.36) & $-$0.05 & $-$0.03 & $-$0.13\\
Step width & 0.21 & 0.19 & 0.19 & 0.04 & 0.24 & 0.21 & weight ($+$0.40) & 0.11 & $-$0.03 & $-$0.09\\
Foot clearance & 0.20 & 0.18 & 0.03 & 0.03 & $-$0.00 & 0.15 & BMI ($-$0.34) & 0.13 & $-$0.12 & $-$0.30\\
Pelvis bob & $-$0.09 & $-$0.09 & 0.05 & 0.01 & $-$0.14 & $-$0.16 & arm/leg ($-$0.50) & 0.18 & 0.16 & 0.06\\
Pelvis sway & $-$0.24 & $-$0.10 & 0.02 & $-$0.06 & $-$0.12 & 0.02 & height ($+$0.24) & $-$0.00 & 0.05 & $-$0.03\\
Pelvis height$^\dagger$ & 0.45 & 0.36 & 0.48 & 0.46 & -- & -- & leg ($-$0.69) & $-$0.28 & $-$0.25 & $-$0.23\\
Knee range & 0.05 & $-$0.00 & 0.13 & 0.11 & $-$0.15 & 0.16 & BMI ($-$0.29) & $-$0.03 & $-$0.12 & $-$0.20\\
Stance knee & $-$0.15 & $-$0.10 & $-$0.10 & $-$0.11 & $-$0.16 & $-$0.09 & arm/leg ($-$0.33) & $-$0.10 & $-$0.05 & $-$0.10\\
Hip range & 0.14 & $-$0.04 & 0.14 & 0.19 & 0.16 & 0.16 & height ($+$0.46) & 0.03 & $-$0.04 & $-$0.20\\
Arm swing & $-$0.01 & $-$0.05 & $-$0.02 & $-$0.08 & $-$0.11 & $-$0.26 & hip w. ($-$0.35) & 0.03 & 0.17 & $-$0.00\\
Elbow angle & $-$0.08 & $-$0.10 & $-$0.14 & $-$0.14 & $-$0.14 & $-$0.29 & arm/leg ($+$0.19) & $-$0.26 & $-$0.26 & $-$0.28\\
Hand distance & 0.01 & 0.07 & $-$0.20 & $-$0.06 & 0.13 & $-$0.15 & female ($+$0.31) & 0.06 & $-$0.05 & $-$0.09\\
Hand reach & $-$0.00 & 0.06 & 0.06 & 0.11 & $-$0.18 & $-$0.02 & hip w. ($+$0.34) & $-$0.23 & $-$0.22 & $-$0.32\\
Forward lean & 0.15 & 0.22 & 0.34 & 0.29 & $-$0.11 & 0.30 & hip w. ($+$0.54) & $-$0.09 & $-$0.08 & $-$0.10\\
Side lean & $-$0.07 & $-$0.11 & $-$0.02 & $-$0.01 & 0.14 & 0.04 & hip w. ($+$0.55) & $-$0.11 & $-$0.11 & $-$0.13\\
Head sway & $-$0.16 & $-$0.11 & 0.08 & 0.07 & $-$0.24 & 0.19 & height ($+$0.57) & $-$0.15 & $-$0.08 & $-$0.22\\
Head bob & $-$0.15 & $-$0.14 & 0.24 & 0.13 & 0.21 & 0.12 & arm/leg ($-$0.54) & 0.17 & 0.15 & $-$0.02\\
Joint speed & $-$0.07 & $-$0.05 & $-$0.05 & $-$0.09 & $-$0.13 & $-$0.06 & BMI ($-$0.18) & $-$0.05 & 0.04 & $-$0.11\\
Jerk & $-$0.03 & 0.00 & $-$0.10 & $-$0.09 & $-$0.10 & $-$0.14 & arm/leg ($-$0.15) & $-$0.00 & 0.03 & $-$0.06\\
Yaw rate$^\dagger$ & $-$0.10 & $-$0.18 & 0.01 & 0.02 & -- & -- & arm/leg ($-$0.31) & $-$0.04 & 0.05 & $-$0.07\\
\bottomrule
\end{tabular}
\label{tab:sup_body_r2}
\end{table*}
The four tests applied to every candidate are the age control, the partial correlation after age, the sign in every style, and the metadata-matched pairs.
Table~\ref{tab:sup_tests} lists the age control and the partial correlation for every descriptor whose best adult pair of body measures has a positive fit.

\begin{table*}[t]
\centering
\scriptsize
\caption{Age control and partial correlation for the 15 descriptors whose best adult pair of body measures has a positive leave-one-out $R^2$.
Refitting the pair on the 31 adults younger than 53 leaves an $R^2$ of at most 0.02 for speed, step length, foot clearance, pelvis bob and sway, knee range, and joint speed, while forward and side lean keep their hip-width correlation after partialling out age.
The pair is chosen on the same 39 adults, so its $R^2$ is optimistic.}
\begin{tabular}{l l cccc cc}
\toprule
& & \multicolumn{4}{c}{Leave-one-out $R^2$} & \multicolumn{2}{c}{Correlation of the pair}\\
\cmidrule(lr){3-6}\cmidrule(l){7-8}
Descriptor & Body pair & Pair & Age & Pair + age & Pair, under 53 & Raw $r$ & $r$ after age\\
\midrule
Speed & height, BMI & 0.07 & 0.30 & 0.27 & $-$0.10 & $+$0.24, $-$0.21 & $+$0.05, $+$0.02\\
Step length & height, BMI & 0.26 & 0.37 & 0.38 & $-$0.07 & $+$0.32, $-$0.33 & $+$0.14, $-$0.12\\
Foot clearance & height, weight & 0.16 & 0.20 & 0.20 & $-$0.03 & $+$0.21, $-$0.14 & $+$0.05, $-$0.12\\
Pelvis bob & height, arm/leg & 0.13 & 0.23 & 0.29 & 0.02 & $+$0.29, $-$0.50 & $+$0.12, $-$0.43\\
Pelvis sway & height, BMI & 0.06 & 0.19 & 0.15 & $-$0.09 & $+$0.24, $-$0.19 & $+$0.08, $+$0.00\\
Knee range & height, weight & 0.18 & 0.22 & 0.22 & 0.01 & $+$0.28, $-$0.07 & $+$0.12, $-$0.03\\
Hip range & height, weight & 0.27 & 0.17 & 0.26 & 0.10 & $+$0.46, $+$0.11 & $+$0.36, $+$0.17\\
Head bob & leg, hip w. & 0.25 & 0.18 & 0.24 & 0.01 & $+$0.49, $-$0.17 & $+$0.39, $-$0.06\\
Joint speed & BMI, hip w. & 0.01 & 0.13 & 0.12 & $-$0.09 & $-$0.18, $+$0.02 & $-$0.00, $+$0.15\\
Forward lean & hip w., arm/leg & 0.40 & $-$0.11 & 0.37 & 0.30 & $+$0.54, $-$0.41 & $+$0.57, $-$0.41\\
Side lean & height, hip w. & 0.13 & $-$0.14 & 0.03 & 0.02 & $+$0.28, $+$0.55 & $+$0.35, $+$0.54\\
Head sway & height, female & 0.24 & $-$0.04 & 0.16 & 0.22 & $+$0.57, $-$0.34 & $+$0.53, $-$0.30\\
Step width & shoulder w., arm/leg & 0.21 & $-$0.09 & 0.26 & 0.23 & $+$0.29, $-$0.33 & $+$0.29, $-$0.37\\
Cadence & hip w., shoulder w. & 0.02 & $-$0.07 & 0.08 & $-$0.07 & $+$0.16, $-$0.16 & $+$0.22, $-$0.18\\
Arm swing & weight, shoulder w. & 0.04 & $-$0.09 & 0.05 & $-$0.03 & $-$0.29, $-$0.03 & $-$0.30, $-$0.02\\
\bottomrule
\end{tabular}
\label{tab:sup_tests}
\end{table*}

Table~\ref{tab:sup_pairs} lists the six adult attribute cells that hold the twelve pairs, and Table~\ref{tab:sup_pair_slopes} compares the slope within these pairs with the slope across the adults for every candidate association.

\begin{table}[t]
\centering
\scriptsize
\caption{The twelve metadata-matched adult pairs come from six attribute cells shared by two or three adults of different body.
Each performer is listed with height (cm) and leg length (m); the two child cells, with two more pairs, are left out.}
\setlength{\tabcolsep}{4pt}
\begin{tabular}{l >{\raggedright\arraybackslash}p{4.3cm} c}
\toprule
Role / affiliation / dominance & Performers & Pairs\\
\midrule
Athlete/sociable/assertive & P12 (189/0.88), P22 (180/0.85), P38 (168/0.74) & 3\\
Professional/moderate/dominant & P34 (160/0.79), P39 (160/0.72) & 1\\
Retiree/moderate/compliant & P01 (159/0.74), P33 (173/0.73) & 1\\
Student/gregarious/assertive & P15 (182/0.85), P32 (175/0.75) & 1\\
Student/reserved/moderate & P28 (160/0.69), P29 (163/0.73), P36 (171/0.80) & 3\\
Student/withdrawn/submissive & P03 (165/0.73), P16 (179/0.80), P31 (164/0.72) & 3\\
\bottomrule
\end{tabular}
\label{tab:sup_pairs}
\end{table}

\begin{table*}[t]
\centering
\scriptsize
\caption{Within the metadata-matched adult pairs, only forward lean (on hip width, BMI, and weight) and side lean (on hip width) keep a rank correlation with $p<0.05$.
The adult slope is fitted across the 39 adults, the paired slope through the origin on the within-pair differences with a bootstrap 95\% interval, in descriptor units per unit of the body measure (hip width and leg length in metres, height in cm, weight in kg).
Agree is the fraction of pairs whose difference has the sign of the adult slope; $\rho$ is the Spearman correlation of the pair differences.}
\begin{tabular}{l l c c c c c c c}
\toprule
Descriptor & Body measure & Pairs & Adult slope & Paired slope & 95\% interval & Agree & $\rho$ & $p$\\
\midrule
Forward lean & hip w. & 12 & $+$226 & $+$131 & [$+$41.3, $+$249] & 0.67 & 0.71 & 0.01\\
Forward lean & BMI & 12 & $+$0.48 & $+$0.177 & [$-$0.0592, $+$0.612] & 0.58 & 0.59 & 0.04\\
Forward lean & weight & 12 & $+$0.0966 & $+$0.0721 & [$+$0.00871, $+$0.193] & 0.67 & 0.72 & 0.01\\
Step length & BMI & 12 & $-$0.00855 & $+$0.00161 & [$-$0.00598, $+$0.00883] & 0.50 & $-$0.06 & 0.85\\
Step length & weight & 12 & $-$0.000453 & $+$0.000497 & [$-$0.00114, $+$0.002] & 0.58 & 0.13 & 0.68\\
Knee range & BMI & 12 & $-$0.679 & $-$0.117 & [$-$1.52, $+$0.686] & 0.50 & $-$0.29 & 0.35\\
Hip range & height & 11 & $+$0.289 & $+$0.277 & [$-$0.00122, $+$0.613] & 0.73 & 0.51 & 0.11\\
Head bob & arm/leg & 12 & $-$0.403 & $-$0.222 & [$-$0.494, $+$0.027] & 0.67 & $-$0.30 & 0.34\\
Step width & weight & 12 & $+$0.00862 & $+$0.00683 & [$+$0.00142, $+$0.0121] & 0.83 & 0.50 & 0.10\\
Foot clearance & BMI & 12 & $-$0.0029 & $-$0.00116 & [$-$0.00662, $+$0.00253] & 0.83 & $-$0.38 & 0.22\\
Side lean & hip w. & 12 & $+$58.8 & $+$73.3 & [$+$18.7, $+$135] & 0.83 & 0.86 & 0.00\\
Head sway & height & 11 & $+$0.000572 & $+$0.000639 & [$+9.66{\times}10^{-5}$, $+$0.00135] & 0.73 & 0.58 & 0.06\\
Hand reach & hip w. & 12 & $+$1.42 & $-$0.013 & [$-$1.81, $+$0.62] & 0.50 & $-$0.10 & 0.76\\
Arm swing & hip w. & 12 & $-$872 & $-$711 & [$-1.88{\times}10^{3}$, $+$358] & 0.50 & $-$0.43 & 0.16\\
Cadence & leg & 12 & $+$0.0457 & $-$0.0509 & [$-$2.26, $+$2.18] & 0.58 & $-$0.21 & 0.51\\
Speed & BMI & 12 & $-$0.0164 & $+$0.016 & [$-$0.00624, $+$0.0405] & 0.50 & 0.15 & 0.63\\
Joint speed & BMI & 12 & $-$0.00548 & $+$0.0073 & [$-$0.00391, $+$0.023] & 0.50 & 0.24 & 0.44\\
Pelvis bob & arm/leg & 12 & $-$0.372 & $-$0.167 & [$-$0.47, $+$0.0629] & 0.67 & $-$0.32 & 0.31\\
\bottomrule
\end{tabular}
\label{tab:sup_pair_slopes}
\end{table*}

\begin{figure}[t]
  \centering
  \includegraphics[width=\columnwidth]{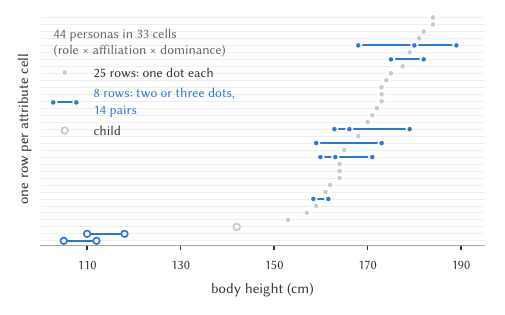}
  \caption{The 44 annotated performers grouped into their 33 role, affiliation, and dominance cells and plotted at their body heights.
  Twenty-five cells hold one performer; the other eight hold two or three, 14 pairs in all, which provide the metadata-matched identities of the real-pair check, not one persona in several bodies.}
  \label{fig:sup_persona_cells}
\end{figure}

\subsection{Age}
Among the adults, age predicts speed, step length, foot clearance, pelvis bob and sway, knee range, and joint speed better than the best pair of body measures (Table~\ref{tab:sup_tests}).
Table~\ref{tab:sup_age} lists the age association, the effect of twenty years, and the comparison of the eight performers aged 53 and over with the 31 younger adults.

\begin{table}[t]
\centering
\scriptsize
\caption{Age lowers speed, step length, foot clearance, pelvis bob and sway, knee and hip range, head bob, and joint speed among the adults, and the effect is carried by the eight performers aged 53 and over.
Effects per +20 years are in the units of Table~\ref{tab:sup_descriptors}; $d$ and the Welch $p$ compare the eight with the 31 younger adults.
The last two columns are the leave-one-out $R^2$ of the age-53 flag and of age within the younger adults alone.}
\setlength{\tabcolsep}{2.6pt}
\begin{tabular}{l ccc c cc cc}
\toprule
& \multicolumn{3}{c}{Age, 39 adults} & & \multicolumn{2}{c}{53 and over} & \multicolumn{2}{c}{$R^2$}\\
\cmidrule(lr){2-4}\cmidrule(lr){6-7}\cmidrule(l){8-9}
Descriptor & $r$ & $R^2$ & $p$ & +20 years & $d$ & $p$ & Flag & Under 53\\
\midrule
Cadence & $-$0.20 & $-$0.07 & 0.34 & $-$0.102 & $-$0.41 & 0.27 & $-$0.07 & $-$0.08\\
Speed & $-$0.61 & 0.30 & 0.00 & $-$0.192 & $-$1.54 & 0.00 & 0.23 & $-$0.00\\
Step length & $-$0.65 & 0.37 & 0.00 & $-$0.0687 & $-$1.78 & 0.00 & 0.30 & $-$0.03\\
Contact ratio & 0.22 & $-$0.10 & 0.62 & $+$0.0235 & 0.57 & 0.46 & -- & --\\
Double support & 0.35 & 0.00 & 0.05 & $+$0.0576 & 0.98 & 0.13 & -- & --\\
Step regularity & 0.04 & $-$0.12 & 0.84 & $+$0.00266 & 0.08 & 0.85 & -- & --\\
Step width & 0.06 & $-$0.09 & 0.51 & $+$0.0202 & 0.13 & 0.70 & $-$0.09 & $-$0.11\\
Foot clearance & $-$0.52 & 0.20 & 0.00 & $-$0.0181 & $-$1.64 & 0.00 & 0.29 & $-$0.18\\
Pelvis bob & $-$0.57 & 0.23 & 0.00 & $-$0.00713 & $-$2.23 & 0.01 & 0.27 & $-$0.13\\
Pelvis sway & $-$0.51 & 0.19 & 0.00 & $-$0.0216 & $-$1.12 & 0.00 & 0.14 & $-$0.03\\
Knee range & $-$0.56 & 0.22 & 0.00 & $-$5.25 & $-$2.04 & 0.00 & 0.32 & $-$0.15\\
Stance knee & $-$0.04 & $-$0.11 & 0.74 & $-$0.175 & $-$0.05 & 0.90 & -- & --\\
Hip range & $-$0.51 & 0.17 & 0.00 & $-$3.52 & $-$1.51 & 0.00 & 0.20 & $-$0.13\\
Arm swing & 0.11 & $-$0.09 & 0.40 & $+$2.97 & 0.55 & 0.07 & $-$0.03 & $-$0.08\\
Elbow angle & 0.10 & $-$0.10 & 0.55 & $+$1.13 & $-$0.09 & 0.79 & -- & --\\
Hand distance & 0.35 & 0.01 & 0.04 & $+$0.0918 & 0.98 & 0.09 & $-$0.00 & $-$0.12\\
Hand reach & 0.22 & $-$0.06 & 0.22 & $+$0.0104 & 0.30 & 0.51 & -- & --\\
Forward lean & $-$0.05 & $-$0.11 & 0.81 & $-$0.245 & $-$0.16 & 0.74 & $-$0.12 & $-$0.22\\
Side lean & 0.12 & $-$0.14 & 0.47 & $+$0.148 & 0.54 & 0.56 & $-$0.18 & $-$0.18\\
Head sway & $-$0.25 & $-$0.04 & 0.13 & $-$0.00278 & $-$0.42 & 0.27 & $-$0.07 & $-$0.08\\
Head bob & $-$0.53 & 0.18 & 0.00 & $-$0.0066 & $-$1.64 & 0.01 & 0.18 & $-$0.11\\
Joint speed & $-$0.48 & 0.13 & 0.00 & $-$0.0591 & $-$1.11 & 0.02 & 0.09 & $-$0.04\\
Jerk & $-$0.36 & 0.05 & 0.02 & $-$36.5 & $-$0.77 & 0.00 & 0.04 & $-$0.07\\
\bottomrule
\end{tabular}
\label{tab:sup_age}
\end{table}

\subsection{Children}
Table~\ref{tab:sup_children} lists the five children against the 39 adults, per child, with the tallest child, P26, separated from the others; the Small column averages three of them, excluding P43's two clips.

\begin{table*}[t]
\centering
\scriptsize
\caption{The children differ from the adults most in foot clearance and step width, but the tallest child is not an interpolation between the small children and the adults.
Difference is children minus adults on the performer effect in the units of Table~\ref{tab:sup_descriptors}, also as a percentage of the grand mean; per-child values are $z$-scores against the adults, with P43's from two clips.
Small is the mean $z$ of P37, P13, and P04, and the last column is P26's $z$ as a fraction of it.}
\begin{tabular}{l ccc cccc c c c}
\toprule
& & & & P43 & P37 & P13 & P04 & & P26 & \\
& Difference & \% of mean & $d$ & 105\,cm, 6\,y & 110\,cm, 5\,y & 112\,cm, 5\,y & 118\,cm, 8\,y & Small & 142\,cm, 9\,y & P26 / small\\
\midrule
Cadence & $+$0.548 & $+$22\% & 1.15 & $+$0.3 & $+$0.5 & $-$0.3 & $+$1.1 & 0.42 & $+$5.1 & 12.24\\
Speed & $+$0.248 & $+$25\% & 0.90 & $+$2.2 & $+$2.5 & $+$0.7 & $+$1.2 & 1.48 & $-$1.8 & $-$1.20\\
Step length & $+$0.0743 & $+$17\% & 0.77 & $+$2.3 & $+$3.0 & $+$0.9 & $+$0.6 & 1.48 & $-$2.3 & $-$1.59\\
Contact ratio & $-$0.0308 & $-$4\% & $-$0.36 & $+$0.4 & $-$1.8 & $-$0.8 & $+$0.1 & -- & $+$0.3 & --\\
Double support & $-$0.0303 & $-$5\% & $-$0.24 & $+$0.4 & $-$1.2 & $-$0.8 & $+$0.1 & -- & $+$0.4 & --\\
Step regularity & $+$0.0641 & $+$14\% & 1.01 & $+$2.9 & $+$2.4 & $+$0.4 & $+$0.4 & -- & $-$0.7 & --\\
Step width & $+$0.442 & $+$22\% & 1.56 & $+$1.7 & $+$2.0 & $+$1.6 & $+$0.8 & 1.46 & $+$1.4 & 0.95\\
Foot clearance & $+$0.135 & $+$68\% & 2.52 & $+$13.6 & $+$6.3 & $+$2.6 & $+$0.6 & 3.13 & $+$1.1 & 0.36\\
Pelvis bob & $+$0.0124 & $+$44\% & 1.27 & $+$1.0 & $+$2.5 & $+$1.3 & $+$0.8 & 1.53 & $+$0.6 & 0.40\\
Pelvis sway & $+$0.02 & $+$23\% & 0.55 & $+$2.1 & $+$1.9 & $+$0.1 & $+$0.2 & -- & $-$1.5 & --\\
Knee range & $+$9.26 & $+$18\% & 1.21 & $+$1.6 & $+$2.7 & $+$1.5 & $-$0.2 & 1.35 & $+$0.4 & 0.32\\
Stance knee & $+$0.136 & $+$0\% & 0.03 & $+$0.9 & $+$1.5 & $+$0.2 & $-$1.5 & -- & $-$0.9 & --\\
Hip range & $+$7.45 & $+$20\% & 1.30 & $+$2.3 & $+$2.7 & $+$1.4 & $-$0.6 & 1.15 & $+$0.9 & 0.79\\
Arm swing & $+$9.17 & $+$10\% & 0.42 & $+$1.1 & $-$0.1 & $+$0.1 & $+$1.2 & -- & $-$0.3 & --\\
Elbow angle & $+$4.55 & $+$6\% & 0.48 & $-$0.0 & $+$1.7 & $+$1.6 & $-$0.2 & -- & $-$0.7 & --\\
Hand distance & $+$0.285 & $+$17\% & 1.34 & $+$1.7 & $+$1.3 & $+$0.1 & $+$3.1 & 1.48 & $+$0.7 & 0.47\\
Hand reach & $-$0.0101 & $-$6\% & $-$0.27 & $+$0.5 & $+$0.2 & $+$0.2 & $-$1.6 & -- & $-$0.7 & --\\
Forward lean & $+$2.03 & $+$39\% & 0.53 & $+$0.8 & $+$2.2 & $-$0.6 & $-$0.5 & -- & $+$0.8 & --\\
Side lean & $+$0.508 & $+$14\% & 0.53 & $+$0.4 & $+$1.6 & $-$0.5 & $-$0.3 & -- & $+$1.4 & --\\
Head sway & $+$0.00513 & $+$10\% & 0.53 & $+$1.4 & $+$2.9 & $-$0.0 & $-$1.7 & -- & $+$0.3 & --\\
Head bob & $+$0.012 & $+$38\% & 1.20 & $+$1.3 & $+$2.8 & $+$1.1 & $+$0.4 & 1.43 & $+$0.4 & 0.30\\
Joint speed & $+$0.11 & $+$21\% & 0.93 & $+$2.9 & $+$3.0 & $+$1.2 & $+$1.0 & 1.76 & $-$2.7 & $-$1.52\\
Jerk & $+$88.5 & $+$49\% & 1.02 & $+$2.0 & $+$2.6 & $+$1.1 & $+$1.3 & 1.68 & $-$1.5 & $-$0.92\\
\bottomrule
\end{tabular}
\label{tab:sup_children}
\end{table*}

\subsection{The two rules in every style}
Table~\ref{tab:sup_lean_styles} lists the slope of forward and side lean on hip width within each of the nine styles, over all styles, and for walking, running, and transitions.

\begin{table}[t]
\centering
\scriptsize
\caption{The slopes of forward and side lean on hip width are positive in every style and in walking, running, and transitions, although several single-style intervals include zero.
Slopes are degrees per +2\,cm of hip width, fitted on the adults' per-performer means, with bootstrap 95\% intervals and leave-one-out $R^2$.}
\setlength{\tabcolsep}{4pt}
\begin{tabular}{l cc cc}
\toprule
& \multicolumn{2}{c}{Forward lean} & \multicolumn{2}{c}{Side lean}\\
\cmidrule(lr){2-3}\cmidrule(l){4-5}
Subset & Slope [95\% interval] & $R^2$ & Slope [95\% interval] & $R^2$\\
\midrule
Angry & $+$3.06 [$+$0.18, $+$6.21] & $+$0.00 & $+$0.43 [$-$0.05, $+$1.12] & $-$0.06\\
Big step & $+$5.11 [$+$1.80, $+$9.34] & $+$0.10 & $+$0.77 [$+$0.29, $+$1.27] & $+$0.07\\
Depressed & $+$9.18 [$+$3.69, $+$15.29] & $+$0.10 & $+$1.21 [$+$0.43, $+$1.99] & $+$0.08\\
Drunk & $+$6.51 [$+$1.05, $+$13.45] & $+$0.04 & $+$1.57 [$+$0.54, $+$2.59] & $+$0.14\\
Fear & $+$10.75 [$+$4.90, $+$18.97] & $+$0.12 & $+$1.01 [$+$0.18, $+$2.18] & $-$0.01\\
Happy & $+$0.55 [$-$1.62, $+$2.44] & $-$0.09 & $+$0.27 [$-$0.19, $+$0.75] & $-$0.07\\
Neutral & $+$2.17 [$-$0.02, $+$4.08] & $+$0.01 & $+$0.29 [$-$0.07, $+$0.73] & $-$0.05\\
Swimming & $+$3.70 [$+$1.01, $+$5.99] & $+$0.08 & $+$0.19 [$-$0.28, $+$0.69] & $-$0.08\\
Two-foot jump & $+$3.54 [$+$0.74, $+$6.23] & $+$0.05 & $+$0.96 [$+$0.11, $+$1.69] & $+$0.05\\
\midrule
All styles & $+$4.69 [$+$2.70, $+$7.40] & $+$0.24 & $+$1.05 [$+$0.42, $+$1.65] & $+$0.17\\
\midrule
Walk & $+$4.81 [$+$2.84, $+$7.60] & $+$0.26 & $+$1.00 [$+$0.38, $+$1.62] & $+$0.14\\
Run & $+$4.67 [$+$2.17, $+$7.65] & $+$0.17 & $+$0.75 [$+$0.36, $+$1.25] & $+$0.07\\
Transitions & $+$4.35 [$+$0.94, $+$8.64] & $+$0.05 & $+$0.69 [$+$0.16, $+$1.27] & $+$0.04\\
\bottomrule
\end{tabular}
\label{tab:sup_lean_styles}
\end{table}

\subsection{What a grid of this size can see}
Table~\ref{tab:sup_power} lists the leave-one-out $R^2$ that a true effect of a given size produces at this sample size.

\begin{table}[t]
\centering
\scriptsize
\caption{At 39 or 44 performers, leave-one-out $R^2$ falls well below the true $R^2$, and a true effect of 0.15 exceeds 0.15 in at most a quarter of the simulations.
Features are Gaussian with one true linear effect; $p=1$ is univariate least squares, $p=4$ ridge on four features of which one is real.
The columns give the median and the 5th and 95th percentiles of the leave-one-out $R^2$, and the probability that it exceeds 0.15 and 0.}
\begin{tabular}{ccc ccc cc}
\toprule
$n$ & $p$ & True $R^2$ & Median & 5\% & 95\% & $P(>0.15)$ & $P(>0)$\\
\midrule
39 & 1 & 0.00 & $-$0.09 & $-$0.14 & 0.02 & 0.00 & 0.07\\
39 & 1 & 0.05 & $-$0.04 & $-$0.12 & 0.14 & 0.05 & 0.32\\
39 & 1 & 0.10 & 0.01 & $-$0.11 & 0.20 & 0.11 & 0.54\\
39 & 1 & 0.15 & 0.06 & $-$0.10 & 0.28 & 0.24 & 0.68\\
39 & 1 & 0.20 & 0.10 & $-$0.06 & 0.31 & 0.40 & 0.84\\
39 & 1 & 0.30 & 0.22 & 0.02 & 0.44 & 0.74 & 0.96\\
39 & 1 & 0.50 & 0.45 & 0.23 & 0.64 & 0.98 & 1.00\\
\midrule
39 & 4 & 0.00 & $-$0.10 & $-$0.18 & $-$0.01 & 0.00 & 0.03\\
39 & 4 & 0.05 & $-$0.08 & $-$0.18 & 0.10 & 0.02 & 0.10\\
39 & 4 & 0.10 & $-$0.03 & $-$0.13 & 0.21 & 0.10 & 0.37\\
39 & 4 & 0.15 & 0.01 & $-$0.11 & 0.29 & 0.15 & 0.57\\
39 & 4 & 0.20 & 0.05 & $-$0.07 & 0.25 & 0.23 & 0.63\\
39 & 4 & 0.30 & 0.16 & $-$0.04 & 0.40 & 0.55 & 0.85\\
39 & 4 & 0.50 & 0.36 & 0.09 & 0.60 & 0.90 & 1.00\\
\midrule
44 & 1 & 0.00 & $-$0.09 & $-$0.12 & 0.00 & 0.00 & 0.05\\
44 & 1 & 0.05 & $-$0.05 & $-$0.11 & 0.14 & 0.05 & 0.26\\
44 & 1 & 0.10 & 0.01 & $-$0.09 & 0.21 & 0.11 & 0.53\\
44 & 1 & 0.15 & 0.07 & $-$0.06 & 0.28 & 0.23 & 0.74\\
44 & 1 & 0.20 & 0.13 & $-$0.03 & 0.33 & 0.41 & 0.89\\
44 & 1 & 0.30 & 0.24 & 0.03 & 0.45 & 0.78 & 0.96\\
44 & 1 & 0.50 & 0.47 & 0.25 & 0.64 & 0.99 & 1.00\\
\midrule
44 & 4 & 0.00 & $-$0.09 & $-$0.17 & $-$0.01 & 0.00 & 0.02\\
44 & 4 & 0.05 & $-$0.06 & $-$0.15 & 0.03 & 0.03 & 0.15\\
44 & 4 & 0.10 & $-$0.04 & $-$0.12 & 0.15 & 0.05 & 0.35\\
44 & 4 & 0.15 & 0.01 & $-$0.07 & 0.19 & 0.10 & 0.52\\
44 & 4 & 0.20 & 0.03 & $-$0.06 & 0.20 & 0.17 & 0.70\\
44 & 4 & 0.30 & 0.18 & 0.01 & 0.43 & 0.65 & 0.97\\
44 & 4 & 0.50 & 0.40 & 0.17 & 0.60 & 0.95 & 1.00\\
\bottomrule
\end{tabular}
\label{tab:sup_power}
\end{table}

\subsection{How much the body can explain}
Let $s_d$ be the share of the total variance of descriptor $d$ carried by the performer effect and $R^2_d$ the best out-of-sample leave-one-out $R^2$ of a predictor, clipped at 0.
Over the $D=23$ motion descriptors, the predictor explains the fractions
\begin{equation}
\eta_{\mathrm{perf}} = \frac{\sum_d s_d R^2_d}{\sum_d s_d}, \qquad \eta_{\mathrm{total}} = \frac{1}{D}\sum_d s_d R^2_d
\end{equation}
of the performer differences and of the total variance, since each descriptor's total variance is normalized to one.
The mean performer share is 0.265, so the second number is about a quarter of the first.
Table~\ref{tab:sup_ceiling} lists the terms and both sums.

\begin{table}[t]
\centering
\scriptsize
\caption{A shape predictor fitted on the adults explains 7.5\% of the performer differences and 2.0\% of the total variance; allowing age nearly doubles both.
Share is the performer share $s_d$, and reliability the Spearman--Brown split-half reliability that bounds any predictor.
The adult columns use shape alone or shape and age; the columns for all 44 use shape with the child flag, shape and age, or the child flag alone.}
\setlength{\tabcolsep}{3pt}
\begin{tabular}{l cc cc ccc}
\toprule
& & & \multicolumn{2}{c}{39 adults} & \multicolumn{3}{c}{All 44}\\
\cmidrule(lr){4-5}\cmidrule(l){6-8}
Descriptor & Share & Reliability & Shape & + age & Shape & + age & Child\\
\midrule
Cadence & 0.21 & 0.84 & 0.00 & 0.00 & 0.00 & 0.00 & 0.00\\
Speed & 0.52 & 0.92 & 0.00 & 0.30 & 0.01 & 0.29 & 0.00\\
Step length & 0.27 & 0.93 & 0.22 & 0.37 & 0.03 & 0.28 & 0.00\\
Contact ratio & 0.33 & 0.93 & 0.07 & 0.07 & 0.04 & 0.04 & 0.00\\
Double support & 0.28 & 0.93 & 0.00 & 0.00 & 0.03 & 0.03 & 0.00\\
Step regularity & 0.12 & 0.80 & 0.01 & 0.01 & 0.00 & 0.00 & 0.00\\
Step width & 0.09 & 0.91 & 0.21 & 0.21 & 0.24 & 0.24 & 0.15\\
Foot clearance & 0.28 & 0.90 & 0.15 & 0.20 & 0.16 & 0.16 & 0.14\\
Pelvis bob & 0.17 & 0.85 & 0.09 & 0.23 & 0.08 & 0.34 & 0.08\\
Pelvis sway & 0.47 & 0.92 & 0.02 & 0.19 & 0.00 & 0.17 & 0.00\\
Knee range & 0.30 & 0.92 & 0.16 & 0.22 & 0.09 & 0.30 & 0.03\\
Stance knee & 0.28 & 0.91 & 0.02 & 0.02 & 0.00 & 0.00 & 0.00\\
Hip range & 0.24 & 0.92 & 0.16 & 0.17 & 0.16 & 0.26 & 0.04\\
Arm swing & 0.19 & 0.77 & 0.03 & 0.03 & 0.04 & 0.04 & 0.00\\
Elbow angle & 0.10 & 0.73 & 0.00 & 0.00 & 0.01 & 0.01 & 0.00\\
Hand distance & 0.27 & 0.88 & 0.00 & 0.01 & 0.13 & 0.13 & 0.06\\
Hand reach & 0.16 & 0.75 & 0.02 & 0.02 & 0.00 & 0.00 & 0.00\\
Forward lean & 0.24 & 0.84 & 0.30 & 0.30 & 0.00 & 0.00 & 0.00\\
Side lean & 0.27 & 0.87 & 0.15 & 0.15 & 0.14 & 0.14 & 0.00\\
Head sway & 0.19 & 0.78 & 0.26 & 0.26 & 0.00 & 0.00 & 0.00\\
Head bob & 0.17 & 0.80 & 0.17 & 0.18 & 0.21 & 0.28 & 0.05\\
Joint speed & 0.41 & 0.86 & 0.00 & 0.13 & 0.06 & 0.14 & 0.00\\
Jerk & 0.51 & 0.86 & 0.00 & 0.05 & 0.06 & 0.12 & 0.00\\
\midrule
Of performer differences & & & 0.075 & 0.145 & 0.058 & 0.140 & 0.018\\
Of total variance & 0.265 & & 0.020 & 0.038 & 0.015 & 0.037 & \\
\bottomrule
\end{tabular}
\label{tab:sup_ceiling}
\end{table}

\section{The retargeting objective in full}
\label{sec:sup_terms}

Table~\ref{tab:sup_terms} lists every term of the retargeting objective summarized in the main paper, what in the body each depends on, and the weights of the production run.

\paragraph{The body}
A target body is a shape vector $\shapevec$, and everything the optimizer needs follows from it.
The SMPL-X joint regressor gives the skeleton for forward kinematics.
The rest-pose mesh, cut by skinning weights into fifteen rigid blocks (pelvis with lower spine, chest, head, and left and right thigh, shank, foot, upper arm, forearm, and hand), gives a signed-distance field per block.
The volume of each block gives its mass and the centre of mass.
The foot vertices give the sole height, four support corners, and the foot length.

\paragraph{The starting point}
The optimizer starts from what geometric retargeting would deliver.
Let the source clip be the local rotations $q^a_t$ and root positions $\mathbf r^a_t$ on body $a$, and let $b$ be the target body.
We copy the rotations, $q^{init}_t = q^a_t$.
We scale the root path by the ratio of leg lengths $r_{leg}$ and the root height by the ratio of hip heights $r_h$:
\begin{equation}
\mathbf r^{init}_t = \mathbf r^a_0 \odot (1, r_h, 1) + (\mathbf r^a_t - \mathbf r^a_0) \odot (r_{leg}, r_h, r_{leg}).
\label{eq:kin_init}
\end{equation}
We then place the lowest toe on the floor and record, for every contact run in the source, where the copy's foot rests; these positions become the anchors below.
This copy is also our stand-in for the standard remedy.
We compared it with the MotionBuilder retargeting used by earlier controllers: on 132 retargets the poses differ by 0.7\,cm after removing the root, and the foot skate is the same.
The persona lives in this copy.
Every later term is a correction to it, and two fidelity terms pull every correction back toward it.

\paragraph{The variables}
The optimizer corrects the copy with a rotation correction $\delta_t \in \mathbb R^{23\times 3}$ per frame and per joint, in axis-angle form, and a root offset $\boldsymbol\rho_t \in \mathbb R^3$:
\begin{equation}
q_t = q^{init}_t \otimes \exp(\delta_t), \qquad \mathbf r_t = \mathbf r^{init}_t + \boldsymbol\rho_t .
\label{eq:vars}
\end{equation}
Corrections on every frame chatter.
Linear interpolation between control points leaves a visible tremor in the standing leg.
We therefore place $\delta$ and $\boldsymbol\rho$ on cubic B-spline control points $\mathbf c_k$ and $\mathbf d_k$, one every three frames:
\begin{equation}
\delta_t = \sum_k B_k(t)\,\mathbf c_k, \qquad \boldsymbol\rho_t = \sum_k B_k(t)\,\mathbf d_k ,
\label{eq:bspline}
\end{equation}
where $B_k$ is the uniform cubic B-spline basis.
The correction is then twice differentiable, and the tremor falls below that of the source capture.
Forward kinematics on the target skeleton turns $(q_t, \mathbf r_t)$ into joint positions $\mathbf p_t$ and block frames, and every term below is a function of these.

\paragraph{The terms}
Every term is an average over the frames of the clip, in centimetres, degrees, or radians, so that one set of weights serves every clip and every body.
Two fidelity terms pull the correction back toward the copy:
\begin{equation}
\begin{aligned}
E_{fid} &= \frac{1}{T}\sum_t \|\delta_t\|^2, \\
E_{root} &= \frac{1}{T}\sum_t \boldsymbol\rho_t^{\!\top} \mathbf W \boldsymbol\rho_t,
& \mathbf W &= \mathrm{diag}(1, 0.3, 1).
\end{aligned}
\label{eq:fid}
\end{equation}
Four terms keep the feet where the copy put them.
Let $\mathbf f_t$ be the four sole corners of a foot, $h_t$ their heights, $c_t \in \{0,1\}$ the source's contact label for that foot, and $T_c$ the number of contact frames.
Then
\begin{align}
E_{ground} &= \frac{1}{T}\sum_t \sum_{\mathrm{corners}} \max(0, -h_t)^2, \nonumber\\
E_{stance} &= \frac{1}{T_c}\sum_t c_t \big(\min_{\mathrm{corners}} h_t\big)^2, \nonumber\\
E_{slide} &= \frac{1}{T_c}\sum_t c_t\, c_{t-1} \|\mathbf f_t - \mathbf f_{t-1}\|^2, \nonumber\\
E_{anchor} &= \frac{1}{T_c}\sum_t c_t \|\mathbf f_t - \hat{\mathbf f}_t\|^2,
\label{eq:feet}
\end{align}
where $\hat{\mathbf f}_t$ is the mean position of the corners over the contact run in the copy, with its lowest corner placed on the floor.
Balance uses the zero-moment point of the mass distribution.
With centre of mass $\mathbf m_t$, its height $m_{t,y}$, and its horizontal acceleration $\ddot{\mathbf m}_{t,xz}$,
\begin{equation}
\begin{aligned}
\mathbf z_t &= \mathbf m_{t,xz} - \frac{m_{t,y}}{g}\,\ddot{\mathbf m}_{t,xz}, \\
E_{bal} &= \frac{1}{T_s}\sum_{t \in \mathcal S}
\left[\max\left(0,\; d(\mathbf z_t, S_t)
- \tfrac12 L_{foot} - 2\,\mathrm{cm}\right)\right]^2 .
\end{aligned}
\label{eq:balance}
\end{equation}
where $S_t$ is the support region, the segment between the two foot centres in double support and the foot centre in single support, and $d$ is the distance to it.
The sum runs over the $T_s$ frames $\mathcal S$ in which at least one foot is in contact.
Joint limits keep every axis-angle component within the source's own range widened by 0.15\,rad, as a squared hinge.
Two smoothness terms penalize the second differences of $\mathbf p_t$ and of $\delta_t$.
The effort term applies a soft budget to a mass-weighted squared-acceleration proxy relative to the source:
\begin{equation}
\mathcal E = \frac{1}{T}\sum_t \sum_{p} m_p \|\ddot{\mathbf m}_{p,t}\|^2, \qquad
E_{effort} = \Big[\max\big(0,\; \mathcal E / \mathcal E_{src} - \tau\big)\Big]^2 ,
\label{eq:effort}
\end{equation}
summing over the blocks $p$ with masses $m_p$ and centres $\mathbf m_{p,t}$.
The last term is the lean association.
With $\theta^{f}_t$ the forward lean of the trunk, $\theta^{s}_t$ its side lean, and $\Delta w$ the hip-width difference between target and source,
\begin{equation}
E_{lean} = \Big(\overline{\theta^{f}} - \overline{\theta^{f}_{src}} - k_f\, \Delta w\Big)^2 + \Big(\overline{|\theta^{s}|} - \overline{|\theta^{s}_{src}|} - k_s\, \Delta w\Big)^2,
\label{eq:lean}
\end{equation}
where the bar is the mean over the clip.
We inject one slope for all styles, $k_f = 150^\circ$ and $k_s = 65^\circ$ per metre of hip width, chosen inside the measured forward-lean range of 131 to 226$^\circ$ per metre reported in the main paper.
The per-style slopes are tabulated in the measurement section above.
The term acts on the mean over the clip, so the style's own time course of lean survives.

\paragraph{Penetration}
The penetration term uses the block distance fields.
For a surface point $x$ of block $P$ and another block $Q$ with signed distance $\phi_Q$, the penetration depth is $\max(0, -\phi_Q(x))$, and the term penalizes only what exceeds a per-point allowance $a(x)$:
\begin{equation}
E_{pen} = \frac{1}{T}\sum_t \sum_{(P,Q)} \sum_{x \in P} \Big[\max\big(0,\, -\phi_Q(x_t) - a(x)\big)\Big]^2 .
\label{eq:pen}
\end{equation}

\paragraph{Where the data calibrates the soft objectives}
Two allowances decide how far the body-aware corrections may go, and the measurement sets both.
The first is the penetration allowance $a(x)$ of Equation~\ref{eq:pen}.
For every surface point we allow the deepest of a 1\,cm soft-tissue margin, the penetration the same point already shows on the source body in this clip, and its overlap in the target's rest pose:
\begin{equation}
\begin{aligned}
a(x) = \max\Big(&1\,\mathrm{cm},\;
\mathrm{P}_{95}\big[\max(0, -\phi_Q(x^a_t))\big]_t,\\
&\max(0, -\phi_Q(x^b_{rest}))\Big).
\end{aligned}
\label{eq:allow}
\end{equation}
Fitting error, skin contact at the armpit or between the thighs, and the rigid-block approximation then cancel between the two bodies, and the optimizer penalizes only the penetration the new body adds.
Without this, a heavy body pushes the hands 4\,cm further out than the real heavy performers do; under the assignment of Section~4.4, normalized hand distance is instead a preservation target.
The second is the effort budget $\tau$ of $E_{effort}$.
The budget is a soft cap.
It only bites when a body would exceed $\tau$ times the source's mass-weighted squared-acceleration proxy for the corrected motion.
That leaves room for the proxy to increase on a heavier body under the same movement.
A strict budget, $\tau = 1$, made a heavy body pay for its mass by cutting arm swing, knee range, and foot clearance, three descriptors the grid says do not move with the body.
The production run therefore sets $\tau = 1.25$ and lets the fidelity terms hold these descriptors at their source values.

\begin{table*}[t]
\centering
\caption{Terms of the retargeting objective, what in the body each depends on, and the weights of the production run.}
\small
\begin{tabular}{l|p{0.52\textwidth}|l|c}
\toprule
Term & What it asks & Depends on & $w$\\
\midrule
Fidelity (rotation, root) & stay close to the kinematic copy & -- & 1, 0.01\\
Penetration & penalize block overlap beyond an allowance & block distance fields & 1\\
Ground & penalize sole corners below the floor & foot geometry & 1\\
Stance & penalize contact height away from the floor & foot geometry & 1\\
Slide & penalize corner velocity during contact & foot geometry & 5\\
Anchor & penalize drift from the copy's contact-run anchors & foot geometry & 10\\
Balance & penalize a quasi-static ZMP proxy outside the support margin & masses, feet & 0.3\\
Source-relative range & penalize joint axes outside the source's range $\pm$0.15\,rad & -- & 10\\
Smoothness & small second differences of positions and corrections & -- & 0.05, 0.02\\
Acceleration proxy & soft budget at 1.25 times the source's & masses & 1\\
Lean & mean forward and side lean at the injected target & hip width & 0.3\\
\bottomrule
\end{tabular}
\label{tab:sup_terms}
\end{table*}

\section{Retargeting at scale}
\label{sec:sup_qc}

The optimizer is Adam over the control points of batches of (clip, body) pairs on GPUs, 400 iterations per batch with the penetration term evaluated every third frame.
One clip and one body cost about five minutes on a CPU; on twelve GPU shards the full run took about two days.
The result is \name-X: 2,570 source clips (we exclude twenty clips shorter than 90 frames) in each of 128 bodies, 328,960 clips in all, frame-aligned with their sources.
Each clip carries the persona ID and typed attributes of its source and the shape vector of its target.

\section{Implementation details}
\label{sec:sup_impl}

Table~\ref{tab:sup_impl} lists the architecture and the training schedule of the two stages and of the raw-space baseline.
Both stages train on the retargeted corpus with the sampler stratified by performer, left-right mirroring with probability 0.5 applied to the motion and the skeleton together, and the root path read 18 frames beyond the window before smoothing, so that the trajectory condition does not brake at the window's end.
Stochastic repeats vary the sampling seed, not the checkpoint.
The latent statistics $\boldsymbol\mu_z$, $\boldsymbol\sigma_z$ of the prior are computed once, from 256 clips drawn with a fixed seed before the first update, and stored in the checkpoint.
The prior represents a persona with four tokens supplied together: one learned 44-way performer-ID embedding and three typed categorical embeddings for role, affiliation, and dominance.
Age and gender are retained as cohort metadata but are not controller conditions.

\paragraph{Evaluation settings}
\begin{figure}[t]
  \centering
  \includegraphics[width=\columnwidth]{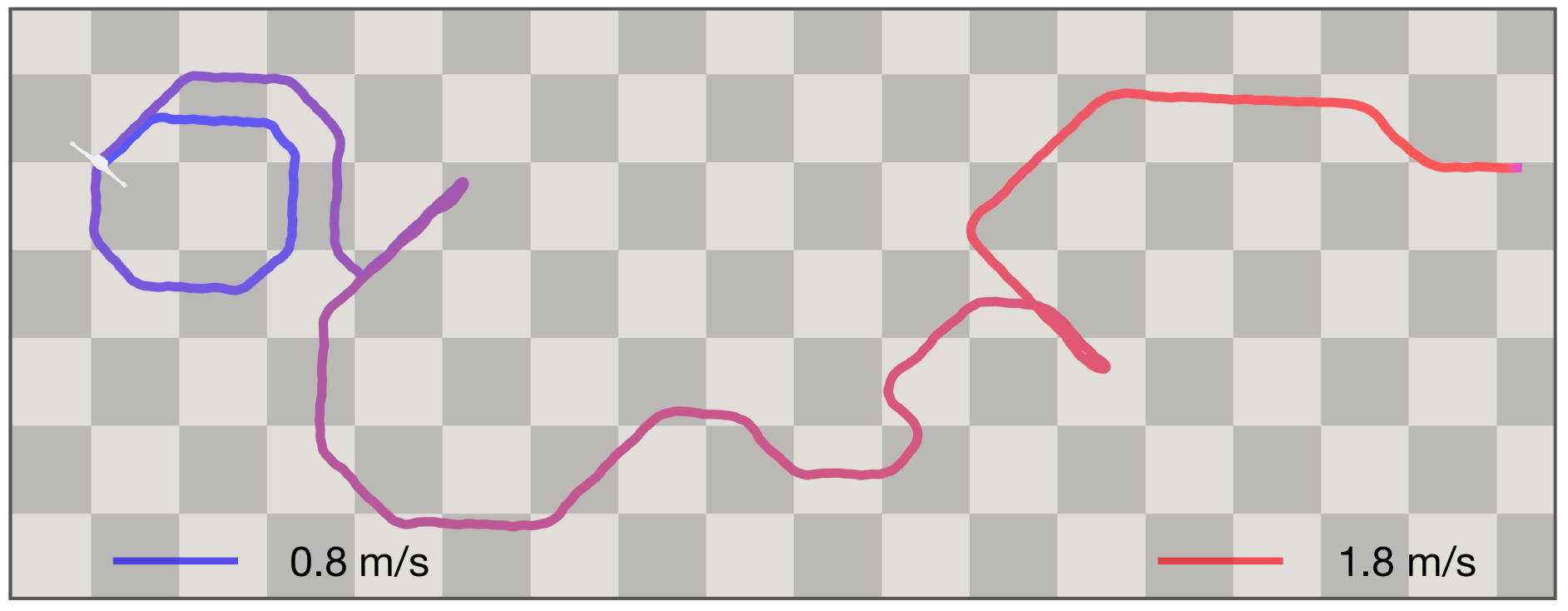}
  \caption{The evaluation trajectory, colored by commanded speed; every test case follows this one-minute control sequence, scaled with leg length.}
  \label{fig:eval_traj}
\end{figure}
The final experiments use the full training partition, with source takes separated before windowing, retargeting, and mirroring.
\paragraph{Partition}
Of the 2,570 source takes, 744 (29\%) are held out: in each persona--style cell with at least four takes, the larger of two and a fifth of them, one take in a cell of three, and none in smaller cells, with the canonical take always kept for training.
The held-out takes of a cell alternate by rank between a reference half (376 takes) and a query half (368 takes).
The fifteen held-out bodies are the grid bodies with index $8i+j \equiv 2 \pmod 5$ over height step $i$ and girth step $j$, excluding index 77.
The sixteen sweep bodies are twelve training bodies at the hip-width quantiles $(2k+1)/24$ and four held-out bodies at the quantiles $(2k+1)/8$.
The withheld cells are the 132 persona--body pairs on the twelve training sweep bodies with $(p + b) \equiv 0 \pmod 4$, where $p$ is the performer's rank in the ID vocabulary and $b$ the body's rank by hip width: three bodies per performer and eleven performers per body.
All rules are fixed with seed 0 before training, and the files' hashes are recorded in every run's configuration.

\paragraph{Rollouts}
The own-body set has 44 personas $\times$ 9 styles, the seen set 44 personas $\times$ the nine training sweep bodies each was paired with, the withheld set the 132 withheld cells, and the held-out set 44 personas $\times$ 15 held-out bodies, each of the last three in all nine styles.
With five sampling seeds, 0 to 4, this gives 1,980, 17,820, 5,940, and 29,700 rollouts per method.
The checkpoint of every model is fixed before any rollout: training has no validation split, and the checkpoint is the one with the lowest training loss after the tenth epoch.

\paragraph{Contacts and aggregation}
A toe is in contact while its speed is below 2\,cm per frame and its height within 5\,cm of the sole plane, smoothed by a three-frame majority vote, with gaps of up to two frames filled and runs shorter than three frames dropped.
Every metric is computed per rollout and averaged over the rollouts of a set; descriptors are averaged over the five seeds of a case before retrieval and spread.

\paragraph{Retrieval and spread}
Persona recovery retrieves, within one body and one style, the nearest of the 44 identities by Euclidean distance after dividing each descriptor by its spread across the references, on six descriptors: jerk, pelvis height, contact ratio, forward lean, hand reach, and motion amplitude.
Style accuracy retrieves the nearest of the performer's nine styles on arm swing, hand reach, head bob, forward lean, and motion amplitude.
Both descriptor sets were chosen by greedy forward selection on the data reference alone, the query half retrieved against the reference half, and never on a model's score.
Spread kept uses joint speed, acceleration, jerk, and high-frequency energy as the dynamics group, and motion amplitude, knee range, foot clearance, pelvis bob, and arm swing as the amplitude group.

\begin{table*}[t]
\centering
\caption{Architecture and training of the two stages and of the raw-space baseline.
Every epoch is about 6.1\,M windows: 6,000 batches of 1,024 split over the GPUs for the two stages, and 2,667 steps of 2,304 windows for the baseline, whose per-GPU batch of 384 with two accumulation steps fits the memory of one card.
The baseline pins its history prefix at every denoising step and therefore trains without history dropout.}
\footnotesize
\setlength{\tabcolsep}{4pt}
\begin{tabular}{@{}p{0.18\textwidth}|p{0.235\textwidth}|p{0.235\textwidth}|p{0.235\textwidth}@{}}
\toprule
 & Shape-aware VAE & Latent FM prior & Raw-space baseline\\
\midrule
Input & 45 frames & 9 tokens $\times$ 48 & 55 frames\\
Token span & 5 frames & -- & --\\
Decoder read-out & 5 frame queries per token; temporal head $3\times$ depthwise conv, kernel 5, residual & -- & --\\
Layers $\times$ width & 6 + 6 $\times$ 384 & 8 $\times$ 384 (DiT) & 8 $\times$ 256\\
Feed-forward, heads & 1024, 4 & 1024, 4 & 1024, 4\\
Dropout & 0.1 & 0.2 & 0.2\\
Conditions & decoder: tail 5 frames, $\shapevec$ & history, trajectory, ID + 3 persona attributes, style, $\shapevec$ & matched conditions; history 10 frames\\
Condition dropout & -- & history token, 15\% & --\\
Sampling & deterministic, $\mathbf z = \boldsymbol\mu$ & 2 Euler steps & 4 DDPM steps\\
Loss & reconstruction stack + KL ($5\times10^{-4}$, free bits 0.1 nats, annealed 20\%) & velocity MSE & same reconstruction stack\\
Epochs $\times$ batches $\times$ batch & 60 $\times$ 6000 $\times$ 1024 & 100 $\times$ 6000 $\times$ 1024 & 60 $\times$ 2667 $\times$ 2304\\
Learning rate & \multicolumn{3}{c}{$10^{-4}$, cosine to $10^{-5}$, weight decay 0.005, EMA 0.995}\\
Parameters & 17.2\,M (8.6 + 8.6) & 19.0\,M & 6.9\,M\\
Training wall clock, RTX 4090 & 11.7\,h on 2 & 18.4\,h on 4 & 24.9\,h on 3\\
Training data & \multicolumn{3}{c}{training split: held-out takes, withheld cells, and held-out bodies excluded}\\
\bottomrule
\end{tabular}
\label{tab:sup_impl}
\end{table*}

\section{Runtime system}
\label{sec:sup_runtime}

\paragraph{Streamer}
The streamer keeps a committed frame stream at 30\,fps and requests a block whenever fewer than three committed frames remain ahead of the playhead.
Each block commits three frames, or twelve when the input has not changed and the character has reached the commanded speed and facing; the tentative remainder of the previous block is cross-faded over two frames with the new one.
A change of input invalidates the queued frames beyond the lead and the block in flight, so that at most one block is ever being computed.

\paragraph{Trajectory and set-points}
The trajectory controller samples the plan at 76 points, the 45-frame window plus a 30-frame lead, integrates the spring-damper response in closed form with sensitivity 8, and anchors the pivot to the true root every frame.
The set-points follow a rule fitted on \name-X: the cruising speed of every clip, defined as the mean over the plateau frames above 0.7 times the clip's 95th percentile, is linear in leg length under leave-one-body-out cross-validation, as the retargeter's scaling implies; the captured grid, by contrast, is best fit by a constant per performer, so the rule is multiplied by a persona factor, each performer's plateau speed relative to the population rule at their leg length.
The factor is capped at the data's own spread, the 90th over the 50th percentile of clip speed (1.52 walking, 1.55 running), because the retargeter carries a child's cadence onto adult legs and four child personas would otherwise ask for a walk above 2\,m/s.
Sideways and backward motion scale the forward set-point by the ratios measured in the data, and a uniform pair of set-points, 1.0\,m/s walking and 1.9\,m/s running for every body, is kept as a switch for comparison.
\paragraph{Display-side post-processing}
Two optional post-processes act on the displayed pose only and never on the committed frames that condition the next block: a foot lock with two-bone leg IK, and a ground estimate taken as the lowest sole over the last 1.5\,s.
The evaluations in the main paper use neither.

\end{document}